\documentclass[nopreprintline,10pt,times]{elsarticle}
\usepackage{amsfonts,amssymb}
\usepackage{amsmath,amsthm}
\usepackage{bm}
\usepackage[caption=false]{subfig}
\usepackage{nomencl}
\makenomenclature
\usepackage{etoolbox}
\renewcommand\nomgroup[1]{%
  \item[\bfseries
  \ifstrequal{#1}{A}{\(\textbf{Sets\ and\ Indices}\)}{%
  \ifstrequal{#1}{C}{\(\textbf{Parameters \ and\ Matrices}\)}{%
  \ifstrequal{#1}{V}{\(\textbf{Variables}\)}{
 }}}%
]}
\journal{Renewable Energy}
\usepackage[noend]{algpseudocode}
\usepackage{algorithm}
\makeatletter
\newcommand{\removelatexerror}{\let\@latex@error\@gobble}
\makeatother
\begin{document}

\begin{frontmatter}



\title{Decentralized Energy Market Integrating Carbon Allowance Trade and Uncertainty Balance in Energy Communities}


\author[inst1]{Yuanxi Wu}

\affiliation[inst1]{organization={School of Electrical Engineering, Southeast University},%Department and Organization
            %addressline={Address One}, 
            city={Nanjing},
            postcode={210096}, 
            %state={Jiang},
            country={China}}

\author[inst1]{Zhi Wu\corref{cor1}}
\author[inst1]{Wei Gu}
\author[inst1]{Zheng Xu}
\author[inst2]{Zheng Shu}
\author[inst1]{Qirun Sun}
\affiliation[inst2]{organization={NARI Technology Co,. Ltd.},%Department and Organization
            %addressline={Address Two}, 
            city={Nanjing},
            postcode={2111062}, 
            %state={State Two},
            country={China}}
\cortext[cor1]{corresponding author: Zhi Wu,  E-mail address: zwu@seu.edu.cn}
\begin{abstract}
With the sustained attention on carbon neutrality, the personal carbon trading (PCT) scheme has been embraced as an auspicious paradigm for scaling down carbon emissions. To facilitate the simultaneous clearance of energy and carbon allowance inside the energy community while hedging against uncertainty, a joint trading framework is proposed in this article. The energy trading is implemented in a peer-to-peer (P2P) manner without the intervention of a central operator, and the uncertainty trading is materialized through procuring reserve of conventional generators and flexibility of users. Under the PCT scheme, carbon allowance is transacted via a sharing mechanism. Possible excessive carbon emissions due to uncertainty balance are tackled by obliging renewable agents to procure sufficient carbon allowances, following the consumption responsibility principle. A two-stage iterative method consisting of tightening McCormick envelope and alternating direction method of multipliers (ADMM) is devised to transform the model into a mixed-integer second-order cone program (MISOCP) and to allow for a fully decentralized market-clearing procedure. Numerical results have validated the effectiveness of the proposed market model.  
\end{abstract}


\end{frontmatter}

%\linenumbers
\section{Introduction}
\label{1}
Traditionally, a large proportion of distribution network load is supplied by centralized fossil-fired power plants, resulting in considerable emissions of greenhouse gas carbon dioxide \cite{RE2}. The ever-worsening climate change has escalated the urgent need of distributed energy resources (DERs) in the distribution network, including micro-turbines (MT),  rooftop photovoltaic (PV) panels and small wind turbines. However, current policies such as feed-in tariff fail to promote the integration of DERs \cite{energyshare} and are insufficient to fulfill the goal of carbon neutrality \cite{RE5}. Recently, the technological advance in energy system management enables a novel electricity market design named peer-to-peer (P2P) energy market \cite{RE1,RE3}, which facilitates the consumption of renewable energy. Decentralized platforms for P2P energy trading transactions with the aid of blockchain technology are developed in \cite{blockchain1,blockchain2}. Besides, generalized Nash game formulation is widely adopted to formulate the energy sharing mechanism \cite{GNE,GNE2}. As for the decentralized optimization algorithm to clear the P2P market, the P2P market is designed as a social welfare maximization problem and the alternating direction multiplier method (ADMM) is employed to achieve consensus among market players \cite{ocadmm,admm2,coordinated}. Other approaches include primal-dual gradient method \cite{primaldual}, Relaxed Consensus + Innovation \cite{consensus}, etc.. Within the aforementioned P2P trading frameworks, individual participants are more inclined to trade directly with their counterparts in the energy community \cite{RE4} rather than with the upstream grid. Therefore, the energy community can reduce the energy loss due to the long-distance transmission and is expected to scale down carbon emissions.

Regarding carbon neutrality, researchers have made endeavors to shed light on low-carbon operations in the power industry. A straightforward approach is to consider low-carbon factors by means of objective functions or price signals. In \cite{objectivecarbon,objectivecarbon2}, the goal of minimizing energy cost is combined with the minimization of ${\rm{C}}{{\rm{O}}_{\rm{2}}}$ emissions and the problem is further formulated as a multi-objective optimization program. In contrast, a energy-carbon integrated price is coined in \cite{pricesignal} based on carbon emission flow and further incentives multiple energy systems to operate in a low-carbon mode implicitly, but it ignores the energy sharing among different entities. As a supplement, reference \cite{objectivecarbon3} considers multi-energy sharing among energy communities and incorporates carbon tax policy to curb carbon emissions.

Another alternative is to introduce a carbon transactive market which is similar to the practice in the energy sector. Carbon market usually refers to a cap-and-trade market \cite{zhang2020emission} where all market participants can trade carbon emission allowances and should surrender corresponding proportion of allowances for the ${\rm{C}}{{\rm{O}}_{\rm{2}}}$ emissions. Conventionally, the production responsibility principle \cite{pro_principle} is adopted, which means energy producers should be accountable for carbon emissions. Recent works have combined the P2P energy market with the carbon market based on this accounting method. In \cite{carbon_blockchain}, all microgrids are motivated to form a grand coalition to transact energy and carbon allowances. Nevertheless, market clearance is solved by the distribution system operator (DSO) and individual privacy concerns may occur. A three-layer framework to trade energy and carbon allowances is established in \cite{carbon_blockchain1}. Notwithstanding the decentralized settling procedure, the exchange of carbon allowances is launched in each time slot, which is scarce in practice.

In recent years, personal carbon trading (PCT) has been viewed as a promising scheme targeted at reducing carbon emissions at the individual and household level \cite{PCT}. The difference is that PCT applies the consumption responsibility principle, i.e., consumers are responsible for carbon emissions precipitated by energy usage \cite{pro_principle,objectivecarbon3}. In a PCT scheme, each consumer is assigned with an initial allocation of carbon allowances and can trade with other consumers. A coupled electricity and emission trading market considering end-users' carbon responsibility is introduced in \cite{wang2020carbon}, but the electricity market is centralized and consumers are penalized for excessive carbon emissions instead of exchanging allowances. The carbon allowances trading is proposed in \cite{PCT1}, while the transactive energy trading is omitted and the identities of allowance sellers/buyers are assigned beforehand.

All of the aforementioned references do not tackle the threat of uncertainty, which is imposed by the presence of increasing penetration of renewable energy sources. Existing works have looked into different approaches to compensate for these uncertainties \cite{gonzalez2021electricity,9618645,8789684,9178314}. In \cite{gonzalez2021electricity}, node-to-node balancing participation factors are leveraged to procure reserve of controllable generators to keep the bulk power system balanced. Flexibility of users is exploited to accommodate deviations of renewable energy outputs in the real-time market via a P2P energy sharing mechanism \cite{9618645}. As for the day-ahead P2P market, the uncertainty is traded with conventional generators or end-users in \cite{8789684,9178314}. Nevertheless, the process of balancing uncertainty is possible to induce more carbon emissions (i.e., emissions resulted from upward reserve supplied by conventional generators), which should be addressed in the carbon market.

Summing up the above, the following issues still need to be further addressed: 1) how to establish a day-ahead decentralized market that can trade energy, uncertainty and carbon allowances jointly in the energy community. 2) how to take into account the exceeding carbon emissions incurred by conventional generators' upward reserve. To this end, this paper proposes a novel community-level P2P market which can trade day-ahead energy, uncertainty and carbon allowances simultaneously. The market participants are constituted of three parts, i.e., renewable agents, conventional generators and users. Renewable agents are supposed to compensate for their forecast errors by procuring reserve from conventional generators and flexibility from users. The definition of flexibility in this paper is the same as that in \cite{9618645}, which is the adjustable capacity the demand can provide in the demand response program. Besides, the carbon market is established under the PCT scheme, and the need to predetermine the participants' identities (sellers or buyers) is obviated 
through a carbon allowance sharing mechanism. 
The main contributions of this paper are summarized as follows.

1)	A joint energy, uncertainty and carbon allowance trading market is developed for the energy community. The proposed framework not only enables energy clearing and carbon allowance sharing simultaneously, but also hedges against the uncertainty.

2)   We leverage the consumption responsibility principle and propose that renewable agents are responsible for acquiring sufficient allowances, which effectively covers potential carbon emissions precipitated due to uncertainty balance and ensures the total emissions are within the prescribed limit.

3)	 A fully decentralized optimization method is developed based on a combination of a modified tightening McCormick method and ADMM, ensuring accuracy while excluding privacy concerns.

The remainder of this paper is organized as follows. Section \ref{2} presents the proposed trading framework and market formulations. Section \ref{3} provides the distributed solution techniques. Case studies are conducted in Section \ref{4}. Finally, the conclusions of this paper follow in Section \ref{5}.

\section{Trading Framework and Market Formulation}
\label{2}
\subsection{Trading Framework}
\begin{figure}[!t]
\centering
\includegraphics[width=4in]{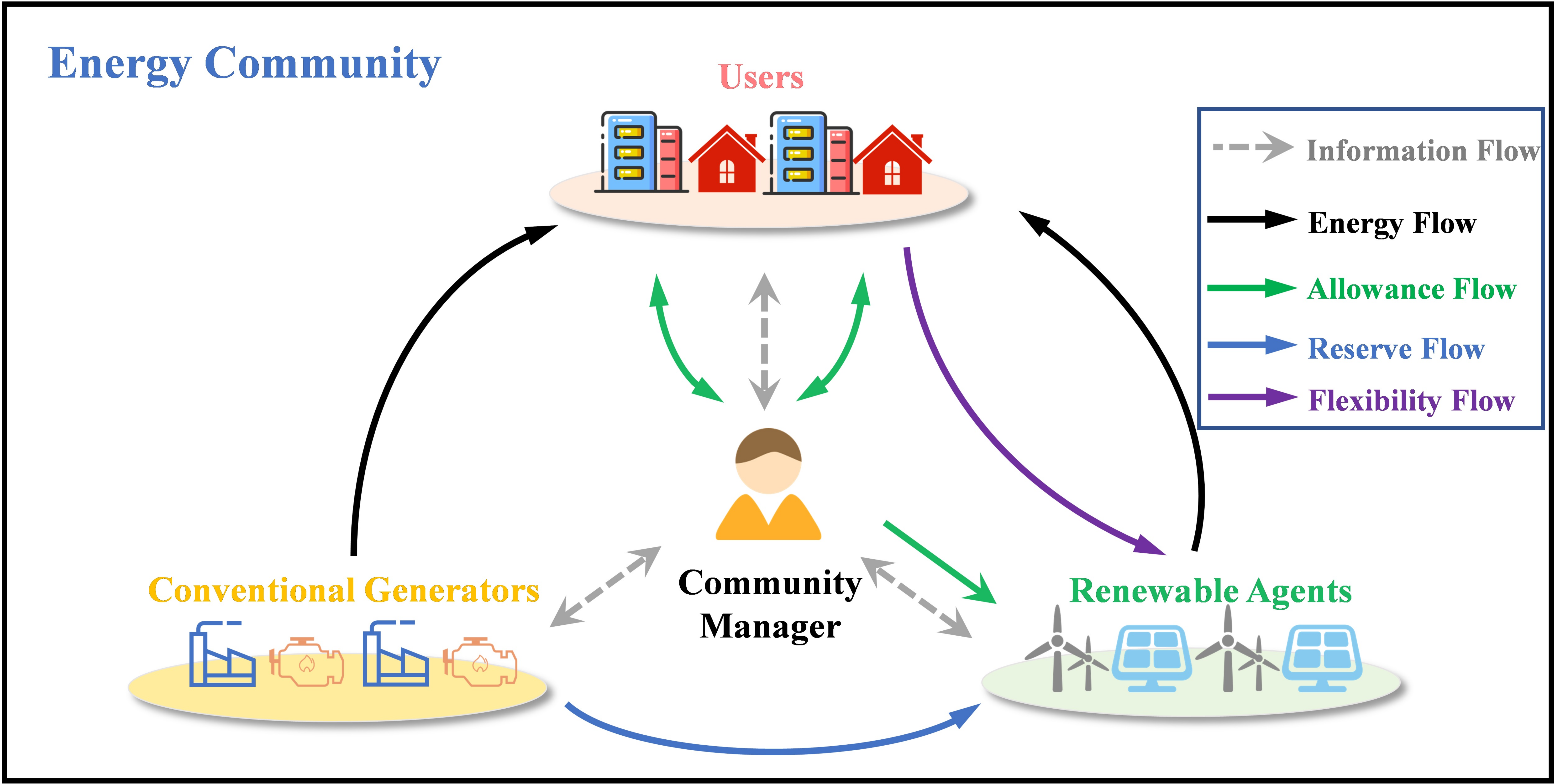}
\caption{Proposed market framework in the energy community}
\label{fig1}
\vspace{-6mm}
\end{figure}

In this paper, a set \(\Omega\) of participants are considered in the joint market, which can be split into three categories, i.e., \(\Omega_{u}\) for users, \(\Omega_{r}\) for renewable agents (RES), such as photovoltaics, and \(\Omega_{g}\) for conventional generators (CG), such as MTs and combined heat and power units (CHP). The joint market is proposed for the day-ahead market and the time interval is 1h. The trading framework is depicted in Fig.  \ref{fig1}.

In the energy market, users choose to buy clean energy from RESs or fossil energy from CGs alternatively. The renewable source generation is featured with uncertainty and thus, RESs need to procure regulating capacity from CGs or users to balance potential forecast errors in the real-time stage, otherwise they will be punished for not fulfilling the contract made in the day-ahead market.

Regarding carbon allowances transactions, users and RESs trade allowances to cover the incurred carbon emissions. Under the PCT scheme, based on individual consumption profiles, users who fail to cover emissions need to purchase allowances in the market, while others with excessive allowances can choose to sell them in the market. The flexibility of users and reserve of CGs contracted in the day-ahead market should be dispatched by RESs who deviate from their predictions at the real-time stage. Then the dispatched reserve becomes another source of carbon emissions. To deal with emissions induced during uncertainty balance, we propose that RESs are accountable and should purchase adequate carbon allowances, which is consistent with the consumption responsibility principle.

Moreover, the users with a surfeit of allowances can sell the allowances to the community manager, which can incentivize users to lead a low-carbon life. The renewable generation not consumed inside the community can be accommodated by the manager as well. All market participants communicate with the community manager to clear the market.
\subsection{Market Formulation}
\subsubsection{Modelling Uncertainty}
Firstly, we model the uncertainty in order to quantify forecast errors. Only the energy deficiency case is considered in this paper since the surplus generation can be curtailed or accommodated by the system operator in the real-time stage \cite{9178314}.

Instead of assuming Gaussian distributed forecast errors, here we only adopt mean and standard deviation of the error to capture uncertainty.
Let \(\boldsymbol{\omega_i^{t^0}}\) denote the random forecast error of RES \textit{i}, which can be divided into two parts: negative component denoted as \(\boldsymbol{\omega_i^{t-}}\) and positive component denoted as \(\boldsymbol{\omega_i^{t+}}\), and it is assumed that \(\mathbb{P} (\boldsymbol{\omega_i^{t^0}}\leq0)=0.5\), \(\mathbb{P} (\boldsymbol{\omega_i^{t^0}}\geq0)=0.5\). Next, to model the case that only generation deficiency is considered, a mixed random variable is defined as follows:
\begin{equation} 
{\boldsymbol{\omega_i^t}} =
\begin{cases}
0,&{\text{if}}\ {\boldsymbol{{\omega}_i^{t^0}}\geq0} ,\\
{\boldsymbol{\omega_i^{t-}},}&{\text{if}}\ {\boldsymbol{\omega_i^{t^0}}\leq0}.
\end{cases}
\end{equation}
Thus, it can be easily deduced that \(\mathbb{E}(\boldsymbol{\omega_i^t})=\frac{1}{2}\mu_i^t\), \(\text{Var}(\boldsymbol{\omega_i^t})=\frac{1}{2}(\delta_i^t)^2+\frac{1}{4}(\mu_i^t)^2\).
\subsubsection{Energy Trading}
The proposed energy market is a bilateral trading market where each participant decides its trading quantity with its neighborhoods. The market equilibrium is represented by the following balancing constraints:
\begin{equation}
Es_{ij}^t+Eb_{ij}^t=0, \quad \forall i\in\Omega_u,j\in\Omega_g\cup\Omega_r
\end{equation}
Each user determines the row vector \(\textbf{Eb}_{i[\cdot]}^t\), while each RES/CG determines the column vector \(\textbf{Es}_{[\cdot] i}^t\). Besides, the trade quantities of sellers are restricted to be non-negative:
\begin{equation}
\textbf{Es}^t\succeq \textbf{0}
\end{equation}
\subsubsection{Uncertainty Trading}
In this paper, the participation factor is adopted to model the bilateral uncertainty trading: $\alpha_{ij}^t$ denotes the participation factor based on which CG \textit{i} produces energy to compensate the uncertainty  $\boldsymbol{\omega_j^t}$, and $\beta_{ij}^t$ denotes the participation factor based on which user \textit{i} is willing to curtail its flexible load to compensate $\boldsymbol{\omega_j^t}$. RESs and CGs, as well as users, should achieve consensus on these uncertainty transactions when reaching the equilibrium:
\begin{equation}
\alpha_{ij}^{r,t}+\alpha_{ij}^t=0, \quad \forall i\in\Omega_g,j\in\Omega_r
\end{equation}
\begin{equation}
\beta_{ij}^{r,t}+\beta_{ij}^t=0,\quad \forall i\in\Omega_u,j\in\Omega_r
\end{equation}
Each RES decides column vectors \(\textbf{A}_{[\cdot]i}^{r,t}\) and \(\textbf{B}_{[\cdot]i}^{r,t}\), while each CG/user decides the row vector \(\textbf{A}_{i[\cdot]}^{t}\)/\(\textbf{B}_{i[\cdot]}^{t}\). Similarly, the participation factors cannot be negative:
\begin{equation}
\textbf{A}^t\succeq\textbf{0}
\end{equation}
\begin{equation}
\textbf{B}^t\succeq\textbf{0}
\end{equation}
RES \textit{j} needs to match the forecast error with the participation factors through uncertainty trading, which means the sum of the participation factors must equal to minus one (since $\alpha_{ij}^{r,t}$/$\beta_{ij}^{r,t}$ and $\alpha_{ij}^{t}$/$\beta_{ij}^{t}$ are opposite in sign):
\begin{equation}
\sum\limits_{i \in \Omega_g} {\alpha_{ij}^{r,t}}+ \sum\limits_{i \in \Omega_u} {\beta_{ij}^{r,t}}=-1,\quad\forall j \in \Omega_r
\end{equation}
\subsubsection{Carbon Market}
As is stated before, the players in the carbon market are users and RESs. The initial daily carbon allowances $\Psi_i^0$ are allocated to users, and then they purchase/sell allowances, respectively, to satisfy individual constraints. Meanwhile, RESs who own no initial allocation have to purchase allowances to counterbalance emissions resulting from upward reserve provided by CGs. Via the sharing mechanism, the carbon allowance trading process can therefore be represented by the balancing constraint below:
\begin{equation}
    \sum\limits_{i\in\Omega_u\cup\Omega_r}c_i=0
\end{equation}

The user who sells allowances in the market can sell them to the community manager alternatively:
\begin{equation}
    0\leq c_i^s\leq M*id_i
\end{equation}
\begin{equation}
    -M*id_i\leq c_i\leq M*(1-id_i)
\end{equation}
where $id_i$ is a binary variable denoting the identity of the user, i.e., 1 for seller while 0 for buyer.

After the clearance of the carbon market, each participant possesses a certain amount of carbon allowances $\Psi_i$:
\begin{equation}
{\Psi_i} =
\begin{cases}
c_i,&{\text{if}}\ \textit{i}\in\Omega_r,\\
\Psi_i^0+c_i-c_i^{s},&{\text{if}}\ \textit{i}\in\Omega_u
\end{cases}
\end{equation}

\noindent
\textbf{\textit{Remark}:} Note that the participants are not permitted to purchase more allowances from the manager since the total initial allocation is set as a cap for the whole community.
\subsubsection{Individual Constraints}
At the equilibrium of the energy market, the power set-point of each participant is equal to the summation of its trade quantity:
\begin{equation}
    p_{u,i}^t=-(\textbf{1})^{\intercal}\cdot\textbf{Eb}_{i[\cdot]}^t,\quad \forall i \in \Omega_u
\end{equation}
\begin{equation}
    p_{r/g,i}^t=\textbf{1}\cdot\textbf{Es}_{[\cdot] i}^t, \quad \forall i \in \Omega_r\cup\Omega_g
\end{equation}
which is also bounded by the following constraints:
\begin{equation}
\underline{p}_{g,i}^t\leq p_{g,i}^t\leq\overline{p}_{g,i}^t,\quad \forall i \in \Omega_g
\end{equation}
\begin{equation}
\underline{p}_{u,i}^t\leq p_{u,i}^t\leq\overline{p}_{u,i}^t,\quad \forall i \in \Omega_u
\end{equation}
\begin{equation}
 p_{r,i}^t+\hat{p}_{r,i}^t= P_{r,i}^t,\quad \forall i \in \Omega_r
\end{equation}
Following (17), we assume that the "green energy" not consumed in the community $\hat{p}_{r,i}^t$ can be accommodated by the community manager.

However, participating in uncertainty balancing induces deviations in the output of CGs and energy consumption of users, which are given by:
\begin{equation}
\widetilde{p}_{g,i}^t=p_{g,i}^t-\boldsymbol{\omega^t}\cdot \textbf{A}_{i[\cdot]}^t, \quad \forall \textit{i} \in \Omega_g
\end{equation}
\begin{equation}
\widetilde{p}_{u,i}^t=p_{u,i}^t+\boldsymbol{\omega^t}\cdot \textbf{B}_{i[\cdot]}^t,\quad\forall i \in \Omega_u
\end{equation}
where \(\boldsymbol{\omega^t}=\begin{pmatrix}\boldsymbol{\omega_1^t }& \boldsymbol{\omega_2^t} & \cdots & \boldsymbol{\omega_{|\Omega_r|}^t}
\end{pmatrix}\) is a random row vector containing all RESs' uncertainties at time \textit{t}, and \(\widetilde{p}_{g,i}^t\)/\(\widetilde{p}_{u,i}^t\) denotes the actual energy set-point of CGs/users, which is a random variable.
Therefore, to ensure the limits are respected, chance-constraints are introduced:
\begin{equation}
\mathbb{P}(\widetilde{p}_{g,i}^t\leq\overline{p}_{g,i}^t)\geq 1-\varepsilon_{g,i},\quad \forall i \in \Omega_g
\end{equation}
\begin{equation}
\mathbb{P}(\widetilde{p}_{u,i}^t\geq\underline{p}_{u,i}^t)\geq 1-\varepsilon_{u,i},\quad \forall i \in \Omega_u
\end{equation}
Constraints (18) and (19) enforce that the power limits should be respected with a predefined probability \(1-\varepsilon_{g/r,i}\), which can be further converted into second-order cone formulations with the aid of Chebyshev approximation\cite{summers2015stochastic}:
 \begin{equation}
 p_{g,i}^t-\textbf{M}^t\cdot\textbf{A}_{i[\cdot]}^t+z_{g,i}S_{g,i}^t\leq\overline{p}_{g,i}^t,\quad \forall i \in \Omega_g
 \end{equation}
 \begin{equation}
 -p_{u,i}^t-\textbf{M}^t\cdot\textbf{B}_{i[\cdot]}^t+z_{u,i}S_{u,i}^t\leq-\overline{p}_{u,i}^t,\quad\forall i \in \Omega_u
 \end{equation}
 where \(\textbf{M}^t=\mathbb{E}(\boldsymbol{\omega^t})\) is the mean value of \(\boldsymbol{\omega^t}\),  and \(z_{g/u,i}=\sqrt{(1-\varepsilon_{g/u,i})/{\varepsilon_{g/u,i}}}\). The covariance matrix of \(\boldsymbol{\omega^t}\) is denoted as \(\boldsymbol{\Sigma}^t\) and the formulations for \(S_{g,i}^t/S_{u,i}^t\) are: \(S_{g,i}^t=\left\lVert (\boldsymbol{\Sigma}^t)^{1/2}(\textbf{A}_{i[\cdot]}^t)^\intercal\right\rVert_2\), \(S_{u,i}^t=\left\lVert (\boldsymbol{\Sigma}^t)^{1/2}(\textbf{B}_{i[\cdot]}^t)^\intercal\right\rVert_2\).
 
 Users' carbon allowances should cover their corresponding emissions:
 \begin{equation}
 CE_i=-\sum\limits_t\sum\limits_{j\in\Omega_g}\sigma_jEb_{ij}^t\leq \Psi_i,\quad \forall i\in \Omega_u
 \end{equation}
 
 While for RESs, potential carbon emissions incurred by dispatching upward reserve can be calculated as follows:
 \begin{equation}
 \widetilde{CE}_i=\sum\limits_t\sum\limits_{j\in\Omega_g}\sigma_j\alpha_{ji}^{r,t}\cdot \boldsymbol{\omega_i^t},\quad\forall i \in \Omega_r
 \end{equation}

RES \textit{i} must procure sufficient allowances to cover the above emissions:
 \begin{equation}
 \mathbb{P}(\widetilde{CE}_i\leq \Psi_i)\geq 1-\varepsilon_{r,i},\quad \forall i \in \Omega_r
 \end{equation}
 Similarly, the above constraint can be transformed into a second-order cone constraint:
 \begin{equation}
 -\mathbb{E}(\boldsymbol{\omega_i})\cdot\textbf{m}_i+\sqrt{(1-\varepsilon_{r,i})/\varepsilon_{r,i}}\left\lVert\boldsymbol{\Xi_i}^{1/2}(\textbf{m}_i)^\intercal\right\rVert_2\leq \Psi_i
 \end{equation}
 \begin{equation}
 m_i^t=-\sum\limits_{j\in\Omega_g}\sigma_j\alpha_{ji}^{r,t},\quad \forall i \in \Omega_r
 \end{equation}
 where \(\boldsymbol{\omega_i}=\begin{pmatrix}
 \boldsymbol{\omega_i^1} & \boldsymbol{\omega_i^2} & \cdots &\boldsymbol{\omega_i^T}
 \end{pmatrix}\) is a row vector containing RES \textit{i}'s uncertainties throughout the scheduling horizon and \(\textbf{m}_i=\begin{pmatrix}
 m_i^1 & m_i^2 & \cdots &m_i^T
 \end{pmatrix}\). \(\boldsymbol{\Xi_i}\) is the covariance matrix of \(\boldsymbol{\omega_i}\).
\subsubsection{Expected Social Welfare Maximization Problem}

It is assumed that all market participants collaboratively minimize the overall cost of the group. Therefore, the objective function can be formulated as follows:
\begin{equation}
\begin{split}
obj=\sum\limits_t[\mathbb{E}(\sum\limits_{i\in\Omega_g}C_i(\widetilde{p}_{g,i}^t))-\mathbb{E}(\sum\limits_{i\in\Omega_u}U_i(\widetilde{p}_{u,i}^t))-\sum\limits_{i\in\Omega_r}r_e^{t}\hat{p}_{r,i}^t]-\sum\limits_{i\in\Omega_u}r_c^sc_i^s
\end{split}
\end{equation}
where \(C_i(p)=c_{2,i}p^2+c_{1,i}p+c_{0,i}\), \(U_i(p)=d_{2,i}p^2+d_{1,i}p\). \(r_e^{t}\) is the selling price for renewable generation at time \textit{t}, and $r_c^s$ is the selling price for carbon allowances. Substituting (18) and (19) into (29), the objective can be further converted into the following expression:

\begin{equation}
\begin{split}
obj & = \sum\limits_t {[\sum\limits_{i \in {\Omega _g}} {({c_{2,i}}{{(p_{g,i}^t)}^2} + {c_{1,i}}p_{g,i}^t + {c_{0,i}}} }  - (2{c_{2,i}}p_{g,i}^t+ {c_{1,i}})(\textbf{M}^t\cdot\textbf{A}_{i[\cdot]}^t) + {c_{2,i}}({(\textbf{M}^t\cdot\textbf{A}_{i[\cdot]}^t)^2} \\
 & + {(S_{g,i}^t)^2})) - \sum\limits_{i \in {\Omega _u}} {({d_{2,i}}{{(p_{u,i}^t)}^2}}+ {d_{1,i}}p_{u,i}^t + (2{d_{2,i}}p_{u,i}^t + {d_{1,i}})(\textbf{M}^t\cdot\textbf{B}_{i[\cdot]}^t)\\
 &+ {d_{2,i}}({(\textbf{M}^t\cdot\textbf{B}_{i[\cdot]}^t)^2}+ {(S_{u,i}^t)^2})) - \sum\limits_{i\in\Omega_r}r_e^{t}\hat{p}_{r,i}^t] -\sum\limits_{i\in\Omega_u}r_{c}^sc_i^s\\
\end{split}
\end{equation}
Summing up the above, the problem can be formulated as:
\begin{equation}
\begin{aligned}
    &\min          &&obj\\
    &\textrm{s.t.} \quad(2)-(17),&& (22)-(24), \quad(27)-(28)
\label{eq:LagrangeRelax}
\end{aligned}
\end{equation}
\section{Distributed Solution Techniques}
\label{3}
In order to solve (31) in a privacy-preserving manner, two obstacles need to be addressed: 1) The uncertainties bring bilinear terms into the objective function (30), which makes it nonconvex function. 2) The constraints (2), (4)-(5) and (9) are coupled among different participants. In this section, we will provide a two-stage iterative method that includes a Relax-ADMM-Contraction loop as described below.
\subsection{Convexification of the Objective Function--- Relax}
The bilinear terms are normally eliminated through McCormick envelopes \cite{mccormick1976computability}. Firstly, the following auxiliary variables are introduced for simplicity:
\begin{equation}
\pi_{g,i}^t=\textbf{M}^t\cdot\textbf{A}_{i[\cdot]}^t,\quad\forall i \in \Omega_g
\end{equation}
\begin{equation}
\pi_{u,i}^t=\textbf{M}^t\cdot\textbf{B}_{i[\cdot]}^t,\quad\forall i \in \Omega_u
\end{equation}
\begin{equation}
\chi_i^t=p_{g,i}^t\pi_{g,i}^t,\quad\forall i \in \Omega_g
\end{equation}
\begin{equation}
\varphi_i^t=p_{u,i}^t\pi_{u,i}^t,\quad\forall i \in \Omega_u
\end{equation}
The lower and upper bounds of $\pi_{g,i}^t$ and $\pi_{u,i}^t$ can be easily deduced as follows:
\begin{equation}
\underline{\pi}_{g,i}^t=\underline{\pi}_{u,i}^t=-\left\lVert\textbf{M}^t\right\rVert_\infty
\end{equation}
\begin{equation}
\overline{\pi}_{g,i}^t=\overline{\pi}_{u,i}^t=0
\end{equation}

Then the McCormick envelope is employed to reformulate the objective as a convex function:
\begin{equation}
\begin{split}
 obj & = \sum\limits_t {[\sum\limits_{i \in {\Omega _g}} {({c_{2,i}}{{(p_{g,i}^t)}^2} + {c_{1.i}}p_{g,i}^t + {c_{0,i}} - 2{c_{2,i}}\chi _i^t} }- {c_{1,i}}\pi _{g,i}^t  \\
 & + {c_{2,i}}({(\pi _{g,i}^t)^2} + {(S_{g,i}^t)^2})) - \sum\limits_{i \in {\Omega _u}} {({d_{2,i}}{{(p_{u,i}^t)}^2}}+ {d_{1,i}}p_{u,i}^t + 2{d_{2,i}}\varphi _i^t  \\
 & + {d_{1,i}}\pi _{u,i}^t + {d_{2,i}}((\pi _{u,i}^t){{\rm{}}^2}+ {(S_{u,i}^t)^2})) - \sum\limits_{i \in {\Omega _r}} {{r_e^{t}}\hat{p}_{r,i}^t}] -\sum\limits_{i\in\Omega_u}r_{c}^sc_i^s
\end{split}
\end{equation}
Additional constraints need to be incorporated:
\begin{subequations}\label{eq:2}
\begin{align}
\chi_i^t&\geq \underline{p}_{g,i}^t\pi_{g,i}^t+\underline{\pi}_{g,i}^tp_{g,i}^t-\underline{p}_{g,i}^t \underline{\pi}_{g,i}\\
\chi_i^t&\geq \overline{p}_{g,i}^t\pi_{g,i}^t+\overline{\pi}_{g,i}^tp_{g,i}^t-\overline{p}_{g,i}^t \overline{\pi}_{g,i}\\
\chi_i^t&\leq \overline{p}_{g,i}^t\pi_{g,i}^t+\underline{\pi}_{g,i}^tp_{g,i}^t-\overline{p}_{g,i}^t \underline{\pi}_{g,i}\\
\chi_i^t&\leq \underline{p}_{g,i}^t\pi_{g,i}^t+\overline{\pi}_{g,i}^tp_{g,i}^t-\underline{p}_{g,i}^t \overline{\pi}_{g,i}\\
\varphi_i^t&\geq \underline{p}_{u,i}^t\pi_{u,i}^t+\underline{\pi}_{u,i}^tp_{u,i}^t-\underline{p}_{u,i}^t \underline{\pi}_{u,i}\\
\varphi_i^t&\geq \overline{p}_{u,i}^t\pi_{u,i}^t+\overline{\pi}_{u,i}^tp_{u,i}^t-\overline{p}_{u,i}^t \overline{\pi}_{u,i}\\
\varphi_i^t&\leq \overline{p}_{u,i}^t\pi_{u,i}^t+\underline{\pi}_{u,i}^tp_{u,i}^t-\overline{p}_{u,i}^t \underline{\pi}_{u,i}\\
\varphi_i^t&\leq \underline{p}_{u,i}^t\pi_{u,i}^t+\overline{\pi}_{u,i}^tp_{u,i}^t-\underline{p}_{u,i}^t \overline{\pi}_{u,i}
\end{align}
\end{subequations}

Following the above procedure, the objective function is transformed into a convex function.
\subsection{Distributed Negotiation Mechanism--- ADMM}
A decentralized market mechanism is essential for keeping transparency and privacy of the joint market and is expected to motivate players in the community to participate. In this paper, a distributed optimization method based on ADMM is adopted to split the global optimization problem into smaller, individual optimization problems. These local problems are solved by market players with limited information exchanges with the community manager. Based on the exchange form of ADMM \cite{boyd2011distributed}, the whole procedure for solving (31) is presented as follows.
\subsubsection{Local Optimization of Each Player}
In the remainder, the cost/utility of each CG/user in (38) will be denoted as $\hat{C}_i^t/\hat{U}_i^t$ for simplicity. 

For each user \textit{i}, its decision variable set is $\xi_i^u=\{\textbf{p}_{u,i}, \textbf{Eb}_{i[\cdot]}, \textbf{B}_{i[\cdot]}, \boldsymbol{\pi}_{u,i}, \boldsymbol{\varphi}_i, c_i, c_i^s, id_i\}$. The local optimization problem of user\textit{ i} at a given iteration \textit{k} is:
\begin{equation}
\setlength{\abovedisplayskip}{3pt}
\setlength{\belowdisplayskip}{3pt}
\begin{alignedat}{2}
   &\xi_i^{u(k+1)}  &&=\arg\min\sum\limits_{t} \big[-\hat{U}_i^t+\sum\limits_{j\in\Omega_r}\lambda_{ij}^{t(k)}\beta_{ij}^t+\sum\limits_{j\in\Omega_r}\frac{\rho}{2}(\beta_{ij}^t-\beta_{ij}^{t(k)}+\hat{\beta}_{ij}^{t(k)})^2\\
   & &&+\sum\limits_{j\in\Omega_r\cup\Omega_g}\upsilon_{ij}^{t(k)}Eb_{ij}^{t}+\sum\limits_{j\in\Omega_r\cup\Omega_g}\frac{\gamma}{2}(Eb_{ij}^{t}-Eb_{ij}^{t(k)}+\hat{E}_{ij}^{t(k)})^2 \big]\\
   & &&+\theta^{(k)}c_i+\frac{\phi}{2}(c_i-c_i^{(k)}+\overline{c}^{(k)})^2-r_{c}^sc_i^s\\
   & \textrm{s.t.} &&(7), (10)-(13), (16), (23)-(24), (33), (39)
    \label{eq:Relax}
\end{alignedat}
\end{equation}

For each RES \textit{i}, its decision variable set is $\xi_i^r=\{\textbf{p}_{r,i}, \textbf{Es}_{[\cdot] i}, \textbf{A}_{[\cdot] i}^r,\textbf{B}_{[\cdot] i}^r, c_i\}$. The local optimization problem of RES \textit{i} at a given iteration \textit{k} is:
\begin{equation}
\begin{alignedat}{2}
&\xi_i^{r(k+1)} &&=\arg\min\sum\limits_{t}\big[-r^{c,t}\hat{p}_{r,i}^t+\sum\limits_{j\in\Omega_u}\lambda_{ji}^{t(k)}\beta_{ji}^{r,t}+\sum\limits_{j\in\Omega_u}\frac{\rho}{2}(\beta_{ji}^{r,t}-\beta_{ji}^{r,t(k)}+\hat{\beta}_{ji}^{t(k)})^2\\
& &&+\sum\limits_{j\in\Omega_g}\eta_{ji}^{t(k)}\alpha_{ji}^{r,t}+\sum\limits_{j\in\Omega_g}\frac{\tau}{2}(\alpha_{ji}^{r,t}-\alpha_{ji}^{r,t(k)}+\hat{\alpha}_{ji}^{t(k)})^2+\sum\limits_{j\in\Omega_u}\upsilon_{ji}^{t(k)}Es_{ji}^t\\
& &&+\sum\limits_{j\in\Omega_u}\frac{\gamma}{2}(Es_{ji}^t-Es_{ji}^{t(k)}+\hat{E}_{ji}^{t(k)})^2\big]+\theta^{(k)}c_i+\frac{\phi}{2}(c_i-c_i^{(k)}+\overline{c}^{(k)})^2\\
& \textrm{s.t.} &&(3), (8), (12), (14), (17), (27)-(28)
\end{alignedat}
\end{equation}

For each CG \textit{i}, its decision variable set is $\xi_i^g=\{\textbf{p}_{g,i},\textbf{Es}_{[\cdot] i},\textbf{A}_{i[\cdot]},\boldsymbol{\pi}_{g,i},\boldsymbol{\chi}_i\}$. The local optimization problem of CG \textit{i} at a given iteration \textit{k} is:
\begin{equation}
    \begin{alignedat}{2}
    &\xi_i^{g(k+1)} &&=\arg\min\sum\limits_t\big[\hat{C}_i^t+\sum\limits_{j\in\Omega_r}\eta_{ij}^{t(k)}\alpha_{ij}^t+\sum\limits_{j\in\Omega_r}\frac{\tau}{2}(\alpha_{ij}^t-\alpha_{ij}^{t(k)}+\hat{\alpha}_{ij}^{t(k)})^2\\
    & &&+\sum\limits_{j\in\Omega_u}\upsilon_{ji}^{t(k)}Es_{ji}^t+\sum\limits_{j\in\Omega_u}\frac{\gamma}{2}(Es_{ji}^t-Es_{ji}^{t(k)}+\hat{E}_{ji}^{t(k)})^2\big]\\
    & \textrm{s.t.} &&(3), (6), (14)-(15), (22), (32), (39)
    \end{alignedat}
\end{equation}
\subsubsection{Global Variable Update}
After gathering all the local information from market players, the community manager is in charge of updating the global variables and then broadcasting the results to all the players. To be specific, the update procedure at a given iteration \textit{k} is as follows:
\begin{subequations}\label{eq:4}
\begin{align}
\hat{\alpha}_{ij}^{t(k+1)}&=\frac{1}{2}(\alpha_{ij}^{t(k+1)}+\alpha_{ij}^{r,t(k+1)})\label{eq:4a}\\
\hat{\beta}_{ij}^{t(k+1)}&=\frac{1}{2}(\beta_{ij}^{t(k+1)}+\beta_{ij}^{r,t(k+1)})\label{eq:4b}\\
\hat{E}_{ij}^{t(k+1)}&=\frac{1}{2}(Eb_{ij}^{t(k+1)}+Es_{ij}^{t(k+1)})\label{eq:4c}\\
\overline{c}^{(k+1)}&=\frac{1}{\vert\Omega_u\cup\Omega_r\vert}\sum\limits_ic_i^{(k+1)}
\end{align}
\end{subequations}
\subsubsection{Dual Price Update}
At the end of each iteration, the dual prices need to be updated following the steps below:
\begin{subequations}\label{eq:5}
\begin{align}
\theta^{(k+1)} &=\theta^{(k)}+\phi\overline{c}^{(k+1)}\label{eq:5A}\\
\lambda_{ij}^{t(k+1)} &=\lambda_{ij}^{t(k)}+\rho\hat{\beta}_{ij}^{t(k+1)} \label{eq:5B}\\
\eta_{ij}^{t(k+1)} &=\eta_{ij}^{t(k)}+\tau\hat{\alpha}_{ij}^{t(k+1)} \label{eq:5D}\\
\upsilon_{ij}^{t(k+1)} &=\upsilon_{ij}^{t(k)}+\gamma\hat{E}_{ij}^{t(k+1)}\label{eq:5G}
\end{align}
\end{subequations}
    
\subsubsection{Stopping Criteria}
The above problem is a convex one except for the nonconvex constraints (10) and (11). Nevertheless, since the non-convexity arises from Boolean constraints and only exists in each user's local problem, the ADMM procedure can still be carried out \cite{9142270,boyd2011distributed}. The proposed distributed mechanism converges as long as the total local residuals fall below the global stopping criteria:
\begin{subequations}\label{6}
\begin{align}
     se^{(k)}&=\sum\limits_{t}\Vert \textbf{Es}^{t(k)}+\textbf{Eb}^{t(k)}\Vert_F^2\leq \epsilon^{pri}_{e}\\
    sr^{(k)}&=\sum\limits_{t}\Vert \textbf{A}^{t(k)}+\textbf{A}^{r,t(k)}\Vert_F^2\leq \epsilon^{pri}_{r}\\
    sd^{(k)}&=\sum\limits_{t}\Vert \textbf{B}^{t(k)}+\textbf{B}^{r,t(k)}\Vert_F^2\leq \epsilon^{pri}_{d}\\
    sc^{(k)}&=(\sum\limits_ic_i^{(k)})^2\leq \epsilon^{pri}_{c}\\
    te^{(k)}&=\sum\limits_{t}\Vert\hat{\textbf{E}}^{t(k)}-\hat{\textbf{E}}^{t(k-1)}\Vert_F^2\leq\epsilon^{dual}_{e}\\
    tr^{(k)}&=\sum\limits_{t}\Vert\hat{\textbf{A}}^{t(k)}-\hat{\textbf{A}}^{t(k-1)}\Vert_F^2\leq\epsilon^{dual}_{r}\\
    td^{(k)}&=\sum\limits_{t}\Vert\hat{\textbf{B}}^{t(k)}-\hat{\textbf{B}}^{t(k-1)}\Vert_F^2\leq\epsilon^{dual}_{d}\\
    tc^{(k)}&=(\overline{c}^{(k)}-\overline{c}^{(k-1)})^2\leq\epsilon^{dual}_{c} 
\end{align}
\end{subequations}
where $\Vert \cdot\Vert_F$ denotes the Frobenius norm, and $\epsilon_e^{pri}\sim\epsilon_c^{dual}$ are the corresponding thresholds.
\subsection{Tightening Bound--- Contraction}
Traditional McCormick envelope usually relax the bilinear term at the sacrifice of accuracy and feasibility. The relaxed version of the market model above renders a lower-bound solution without promising the feasibility of the original model. Hence, a heuristic bound contraction algorithm modified from \cite{deng2021optimal} is adopted in this paper to improve the precision of the traditional McCormick envelopes, which can iteratively strengthen the bounds of $p_{g,i}^t,p_{u,i}^t,\pi_{g,i}^t$ and $\pi_{u,i}^t$. This is achieved by using a decreasing scalar to tighten the bounds according to the solutions from the last iteration. Besides, as stated in \cite{deng2021optimal}, the updated bounds should be the intersection of the result-oriented bounds and the initial bounds to ensure the feasibility of the original model. Therefore, at a given iteration $n$, the bounds should be updated based on the following rules:
\begin{subequations}
\begin{align}
    \underline{p}_{g,i}^{t}&=\max\{(1-\epsilon^n)p_{g,i}^{t*},\underline{p}_{g,i}^{t,ini}\} \\
    \overline{p}_{g,i}^{t}&=\min\{(1+\epsilon^n)p_{g,i}^{t*},\overline{p}_{g,i}^{t,ini}\} \\
    \underline{p}_{u,i}^{t}&=\max\{(1-\epsilon^n)p_{u,i}^{t*},\underline{p}_{u,i}^{t,ini}\} \\
    \overline{p}_{u,i}^{t}&=\min\{(1+\epsilon^n)p_{u,i}^{t*},\overline{p}_{u,i}^{t,ini}\} \\
    \underline{\pi}_{g,i}^{t}&=\max\{(1+\epsilon^n)\pi_{g,i}^{t*},\underline{\pi}_{g,i}^{t,ini}\} \\
    \overline{\pi}_{g,i}^{t}&=\min\{(1-\epsilon^n)\pi_{g,i}^{t*},\overline{\pi}_{g,i}^{t,ini}\} \\
    \underline{\pi}_{u,i}^{t}&=\max\{(1+\epsilon^n)\pi_{u,i}^{t*},\underline{\pi}_{u,i}^{t,ini}\} \\
    \overline{\pi}_{u,i}^{t}&=\min\{(1-\epsilon^n)\pi_{u,i}^{t*},\overline{\pi}_{u,i}^{t,ini}\} 
\end{align}
\end{subequations}
where $\epsilon^n=\epsilon^{n-1}-\kappa$ is a decreasing scalar, $(\cdot)^*$ denotes the solution from the last iteration and $(\cdot)^{ini}$ denotes the bound used in the first iteration. The discrepancy between signs in (48) and (49) is because $\pi_{u,i}^t$ and $\pi_{g,i}^t$ are always non-positive while $p_{u,i}^t$ and $p_{g,i}^t$ are always non-negative.

Upon the updates of the bounds of the decision variables, the McCormick envelopes (39) and (40) are updated accordingly.

The update procedure will terminate once the maximal relative error of the bilinear constraints (34) and (35) fall below a reasonable level:
\begin{subequations}
\begin{align}
    err^g&=\max\limits_{t,i}\vert(\chi_i^t-p_{g,i}^t\pi_{g,i}^t)/\chi_i^t\vert \leq\delta^g\\
    err^u&=\max\limits_{t,i}\vert(\varphi_i^t-p_{u,i}^t\pi_{u,i}^t)/\varphi_i^t\vert\leq\delta^u
\end{align}
\end{subequations}

To sum up, the whole procedure of the Relax-ADMM-Contraction loop is presented in Algorithm 1.
\begin{figure}[!t]
\renewcommand{\algorithmicrequire}{\textbf{Input:}}
  \renewcommand{\algorithmicensure}{\textbf{Output:}}
  \removelatexerror
  \begin{algorithm}[H]
    \caption{Relax-ADMM-Contraction}
    %\algsetup{indent=2em}
    \begin{algorithmic}[1]
    \iffalse
      \ENSURE $\hat{\textbf{E}}^t,\hat{\textbf{A}}^t,\hat{\textbf{B}}^t,c_i,p_{g,i}^t,p_{u,i}^t,p_{r,i}^t$
      \REPEAT \COMMENT{Tightening McCormick Envelope} 
      \STATE Derive the relaxed original problem (38)-(39).
      \STATE \textbf{Initialization:} $k\gets0$, dual prices, global variables
      \REPEAT[ADMM Procedure]
      \STATE \textbf{Local optimization of each player:}
      \FORALL{$i\in\Omega_u$} 
      \STATE Solve (40) and obtain $\xi_i^{u(k+1)}$
      \ENDFOR
      \FORALL{$i\in\Omega_r$}
      \STATE Solve (41) and obtain $\xi_i^{r(k+1)}$
      \ENDFOR
      \FORALL{$i\in\Omega_g$}
      \STATE Solve (42) and obtain $\xi_i^{g(k+1)}$
      \ENDFOR
      \STATE \textbf{Community manager coordination:}
      \STATE Updates and broadcasts global variables: (43)
      \STATE Updates and broadcasts dual prices: (44)
      \STATE $k\gets k+1$
      \UNTIL{convergence conditions (45) is satisfied}
      \STATE \textbf{Bound Contraction:}
      \FORALL{$i\in\Omega_u\cup\Omega_g$}
      \STATE Update local decision variables (46) 
      \STATE Update local constraints (39) 
      \ENDFOR
      \STATE $\epsilon^{n+1}=\epsilon^n-\kappa$
      \STATE $n\gets n+1$
      \UNTIL{convergence condition (47) is satisfied}
      \fi
      \Ensure $\hat{\textbf{E}}^t,\hat{\textbf{A}}^t,\hat{\textbf{B}}^t,c_i,p_{g,i}^t,p_{u,i}^t,p_{r,i}^t$
      \Repeat \Comment{Tightening McCormick Envelope} 
      \State Derive the relaxed original problem (38)-(39).
      \State \textbf{Initialization:} $k\gets0$, dual prices, global variables
      \Repeat \Comment{ADMM Procedure}
      \State \textbf{Local optimization of each player:}
      \ForAll{$i\in\Omega_u$} 
      \State Solve (40) and obtain $\xi_i^{u(k+1)}$
      \EndFor
      \ForAll{$i\in\Omega_r$}
      \State Solve (41) and obtain $\xi_i^{r(k+1)}$
      \EndFor
      \ForAll{$i\in\Omega_g$}
      \State Solve (42) and obtain $\xi_i^{g(k+1)}$
      \EndFor
      \State \textbf{Community manager coordination:}
      \State Updates and broadcasts global variables: (43)
      \State Updates and broadcasts dual prices: (44)
      \State $k\gets k+1$
      \Until{convergence conditions (45) is satisfied}
      \State \textbf{Bound Contraction:}
      \ForAll{$i\in\Omega_u\cup\Omega_g$}
      \State Update local decision variables (46) 
      \State Update local constraints (39) 
      \EndFor
      \State $\epsilon^{n+1}=\epsilon^n-\kappa$
      \State $n\gets n+1$
      \Until{convergence condition (47) is satisfied}
    \end{algorithmic}
  \end{algorithm}
  \vspace{-6mm}
  \end{figure}
  
\noindent
\textbf{\textit{Remark}:} To accelerate the convergence, the solutions of dual prices and global variables from the outer iteration $n-1$ will be adopted as initial values at the iteration $n$. The effectiveness of this warm-start method will be illustrated in Section \ref{4}.

We now state that the dual prices exactly constitute the competitive market equilibrium prices.

\noindent
\textbf{\textit{Proposition 1.}} Using $\pi^{e,t}_{ij}=-\upsilon_{ij}^t, \pi^{u,t}_{ij}=-\eta_{ij}^t, \pi^{d,t}_{ij}=-\lambda_{ij}^t, \pi^c=\theta$ as the bilateral energy prices, upward reserve prices, flexibility prices and carbon allowance price, respectively, constitutes a competitive market equilibrium. 

\noindent
\textbf{\textit{Proof.}} We firstly define the individual profit-maximizing problem for RES \textit{i} as follows:
\begin{equation}
    \begin{alignedat}{2}
    &\min \quad&&\sum\limits_t\big[-r^{e,t}\hat{p}_{r,i}^t-\sum\limits_{j\in\Omega_u}\pi_{ji}^{d,t}\beta_{ji}^{r,t}-\sum\limits_{j\in\Omega_g}\pi_{ji}^{u,t}\alpha_{ji}^{r,t}\\
    & &&-\sum\limits_{j\in\Omega_u}\pi_{ji}^{e,t}Es_{ji}^t\big]+\pi^cc_i\\
    &\rm{s.t.}\quad&&(3), (8), (12), (14), (17), (27)-(28)
    \end{alignedat}
\end{equation}

It can be inferred from (41) that the market outcome will solve the following local optimization problem for RES \textit{i} after the convergence of the market:
\begin{equation}
    \begin{alignedat}{2}
    &\min \quad&&\sum\limits_t\big[-r^{e,t}\hat{p}_{r,i}^t+\sum\limits_{j\in\Omega_u}\lambda_{ji}^{t}\beta_{ji}^{r,t}+\sum\limits_{j\in\Omega_g}\eta_{ji}^{t}\alpha_{ji}^{r,t}+\sum\limits_{j\in\Omega_u}\upsilon_{ji}^{t}Es_{ji}^t\big]+\theta c_i\\
    &\rm{s.t.}\quad&&(3), (8), (12), (14), (17), (27)-(28)
    \end{alignedat}
\end{equation}

Applying $\pi^{e,t}_{ij}=-\upsilon_{ij}^t, \pi^{u,t}_{ij}=-\eta_{ij}^t, \pi^{d,t}_{ij}=-\lambda_{ij}^t, \pi^c=\theta$, it can be found that the outcome of the market will solve the individual profit-maximizing problem (49) likewise. The same procedure can be applied to users and CGs as well. Based on \cite[Definition 1]{o2005efficient}, the set of prices  $\{\pi^{e,t}_{ij},\pi^{u,t}_{ij},\pi^{d,t}_{ij},\pi^c\}$ constitutes a competitive equilibrium.  \qed
\section{Case Study}\label{4}
\subsection{Case Parameters}
In this study, the energy community consisting of eight market participants, i.e., three MTs, three users and two PV generators is presented. The parameters of MTs and users are listed in Table \ref{t1}\&\ref{t2}. The forecast output curves of PVs are depicted in Fig. 2 and the standard deviations $\sigma_i^t$ for the PV prediction errors are set as 10\% of the predicted outputs. Then, the standard deviation and mean value of the negative error component are generated based on the following rules\cite{gonzalez2021electricity}:
\begin{equation}
    \delta_i^t=\sigma_i^t\sqrt{\frac{\pi-2}{\pi}},\quad\mu_i^t=\sigma_i^t\sqrt{\frac{2}{\pi}}
\end{equation}
\begin{table}[htp]
\begin{center}
%\small
\caption{Parameters of micro-turbines}
\label{t1}
\renewcommand{\arraystretch}{1.2}
\begin{tabular}{| c | c | c | c |}
\hline
Parameters & MT1 & MT2 & MT3\\
\hline
$c_0\;\big[\$\big]$ & 2.01 & 2.01 & 2.03\\
\hline
$c_1\;\big[\$/\rm{kW}\big]$ & 0.045 & 0.050 & 0.052\\
\hline
$c_2\;\big[\$/\rm{kW}^2\big]$ & 0.00021 & 0.00021 & 0.00019\\
\hline
$\overline{p}_g\;\big[\rm{kW}\big]$ & 260 & 270 & 220\\
\hline
$\underline{p}_g\;\big[\rm{kW}\big]$ & 0 & 0 & 0\\
\hline
$\sigma_i\;\big[\rm{kg}\cdot \rm{kW}^{-1}\big]$ & 0.870 & 0.935 & 0.910\\
\hline
\end{tabular}
\end{center}
\vspace{-6mm}
\end{table}
\begin{table}[htp]
\begin{center}
%\small
\caption{Parameters of users}
\label{t2}
\renewcommand{\arraystretch}{1.2}
\begin{tabular}{| c | c | c | c |}
\hline
Parameters & U1 & U2 & U3\\
\hline
$d_1\;\big[\$/\rm{kW}\big]$ & 0.0870 & 0.0765 & 0.0600\\
\hline
$d_2\;\big[\$/\rm{kW}^2\big]$ & -0.00014 & -0.00014 & -0.000125\\
\hline
%$\Psi_i^0\;\big[\rm{kg}\big]$ & 1800 & 1800 & 1800\\
%\hline
\end{tabular}
\end{center}
\vspace{-6mm}
\end{table}
\begin{figure}[htp]
\centering
\includegraphics[width=0.8\linewidth]{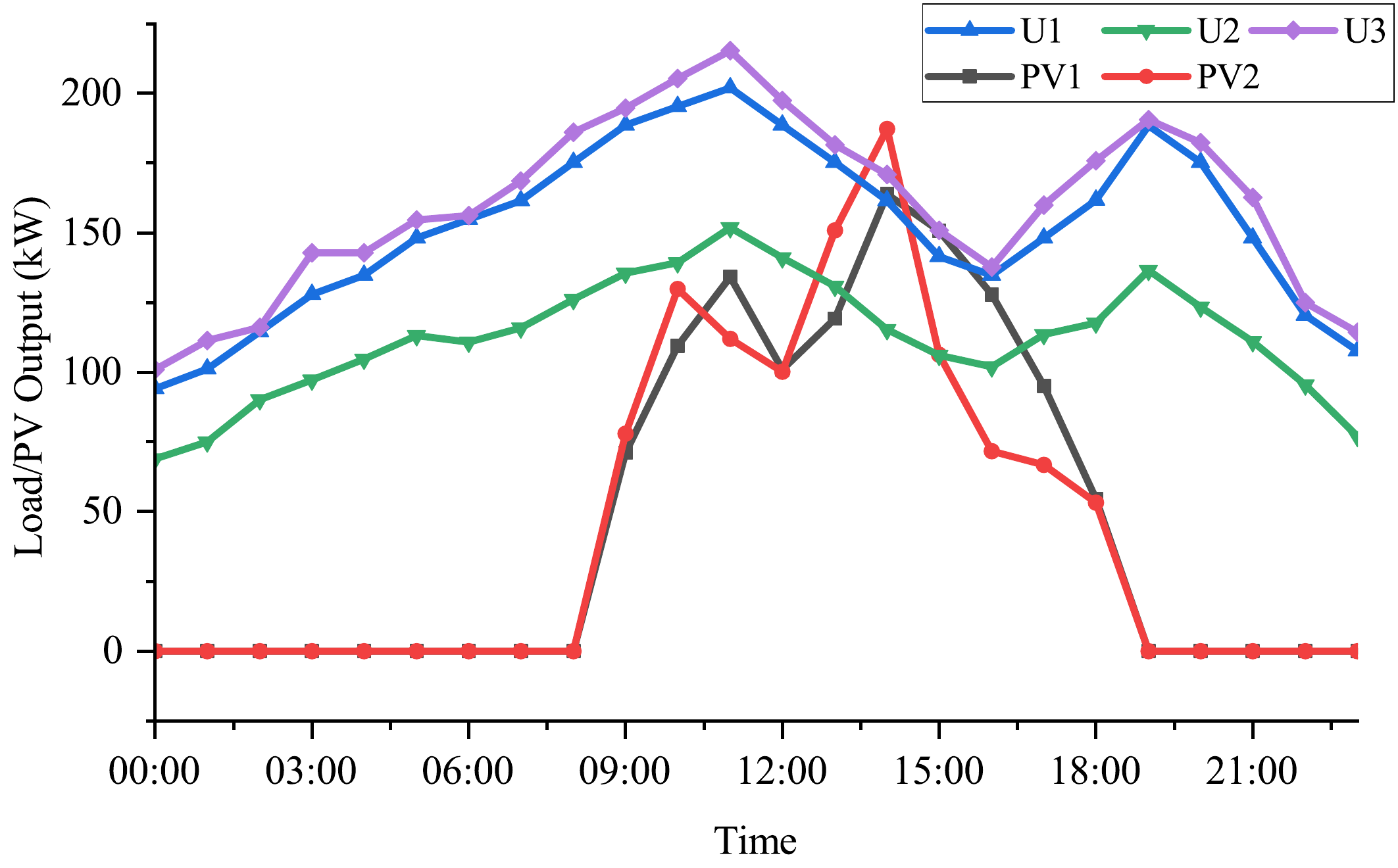}
\caption{Day-ahead PV forecast output and upper bound of loads}
\label{fig2}
\vspace{-3mm}
\end{figure}

The upper bound of the load profile of each user is shown in Fig. 2 and the lower bound is set to 40\% of the upper bound. The selling price for carbon allowances $r_c^s$ is set as \$0.003/kg and the day-ahead selling price for electricity $r_e^t$ is set as \$0.06/kWh. The initial carbon allowance allocation is based on an equal per capita allocation, which is 1800 kg per user in this study. All the confidence levels for chance constraints are set as 0.95. In the ADMM procedure, the tolerance levels for primal residuals are set to $10^{-6}$ for $sr$ and $sd$, and $10^{-4}$ for $se$ and $sc$. Tolerance levels for dual residuals are chosen the same as the corresponding primal residuals. The stopping criteria for the bound contraction are chosen to be $10^{-2}$.
All the optimization problems are carried out in MATLAB 2021b platform using Gurobi \cite{gurobi} solver along with Yalmip\cite{yalmip}. The simulations run on a computer featuring AMD Ryzen 7 5800H @3.20GHz and 16 GB of RAM.
\subsection{Market Outcome}
In this part, we will firstly discuss the outcomes of the proposed joint market. Fig. 3 and Fig. 4 present the outcomes of electricity trading and uncertainty balance, respectively. Fig. 3 illustrates consumption profile for each user, including the load, lower bound of the load (Lb) and components of the load. It can be seen that for each user, the electricity purchase and demand reach the balance in each time slot. In addition, since MT3 is more expensive and has a relatively high carbon intensity, users procure the least energy from MT3 as revealed in Fig. 3. Conversely, MT1, who has the least cost parameter and carbon intensity, possesses the largest percentage among all three MTs. This statement also explains the absence of MT3 in uncertainty balance as depicted in Fig. 4, which illustrates the participation factors of users and MTs in compensating for the uncertainty. It is obvious that for each PV, the sum of the participation factors of users and MTs is equal to 1, which implies that the forecast error can be fully offset. Meanwhile, user3 does not participate in 
uncertainty balance during all time periods because it has reached its lower bound in the energy market as shown in Fig. 3, excluding itself from providing flexibility. 
\begin{figure}[htp]
\centering
\subfloat[User 1]{\includegraphics[width=0.45\textwidth]{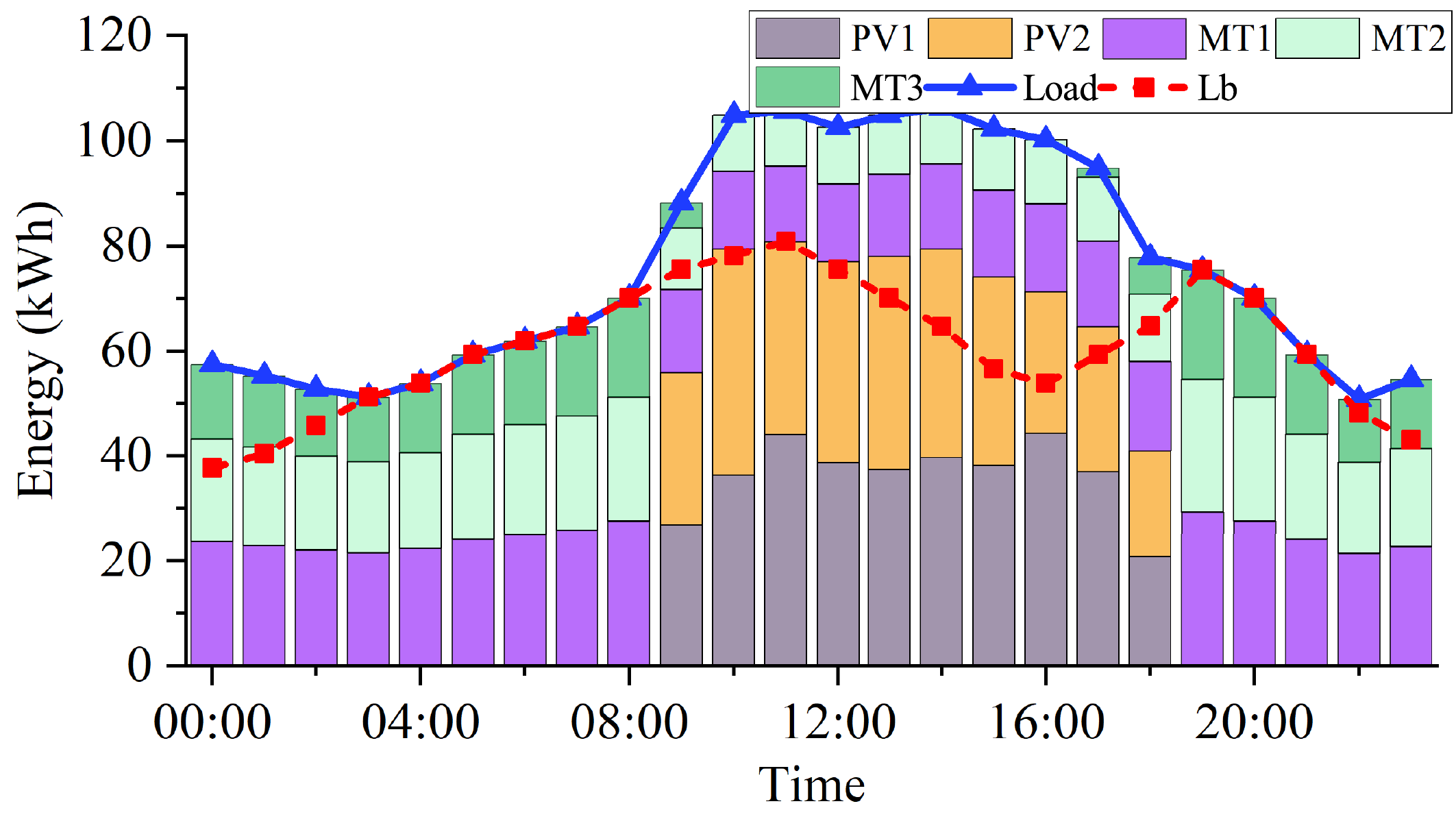}}\hfill
\subfloat[User 2]{\includegraphics[width=0.45\textwidth]{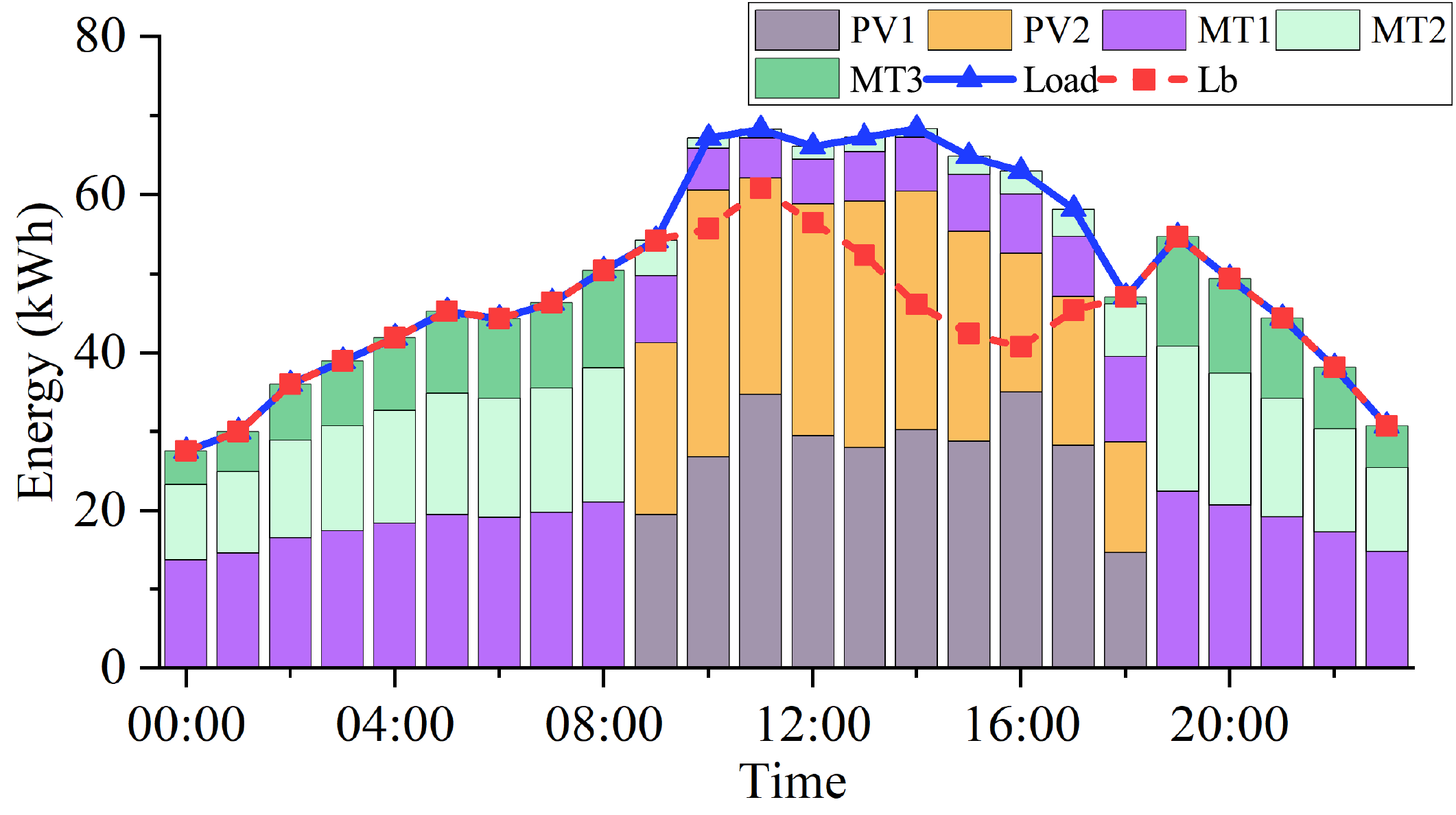}}\par
\subfloat[User 3]{\includegraphics[width=0.45\textwidth]{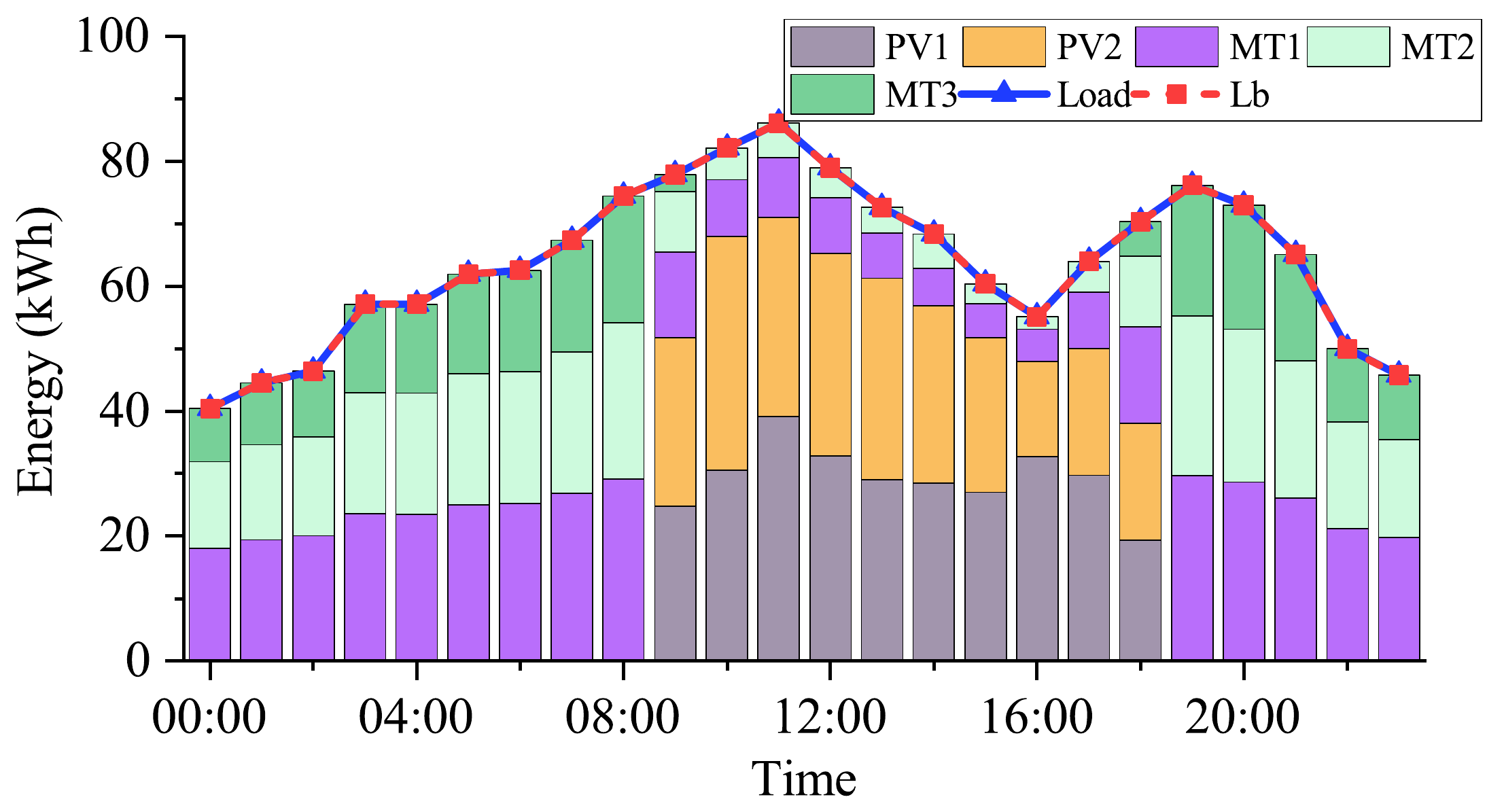}}
\caption{Optimal energy consumption of each user in the day-ahead market}
\label{fig3}
\vspace{-6mm}
\end{figure}
\begin{figure}[htp]
\centering
\subfloat[PV 1]{\includegraphics[width=0.5\textwidth]{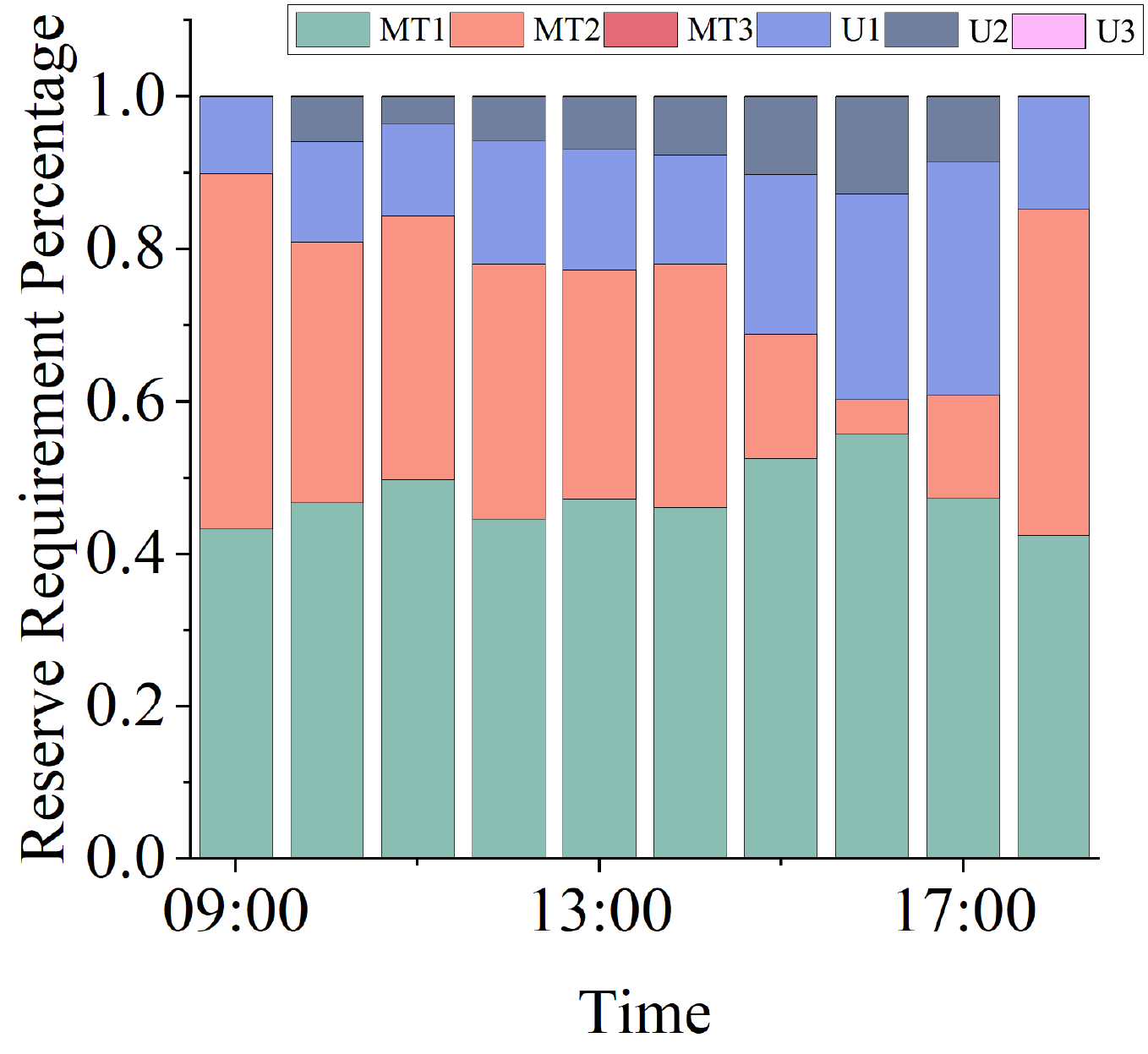}}
%\quad
\subfloat[PV 2]{\includegraphics[width=0.5\textwidth]{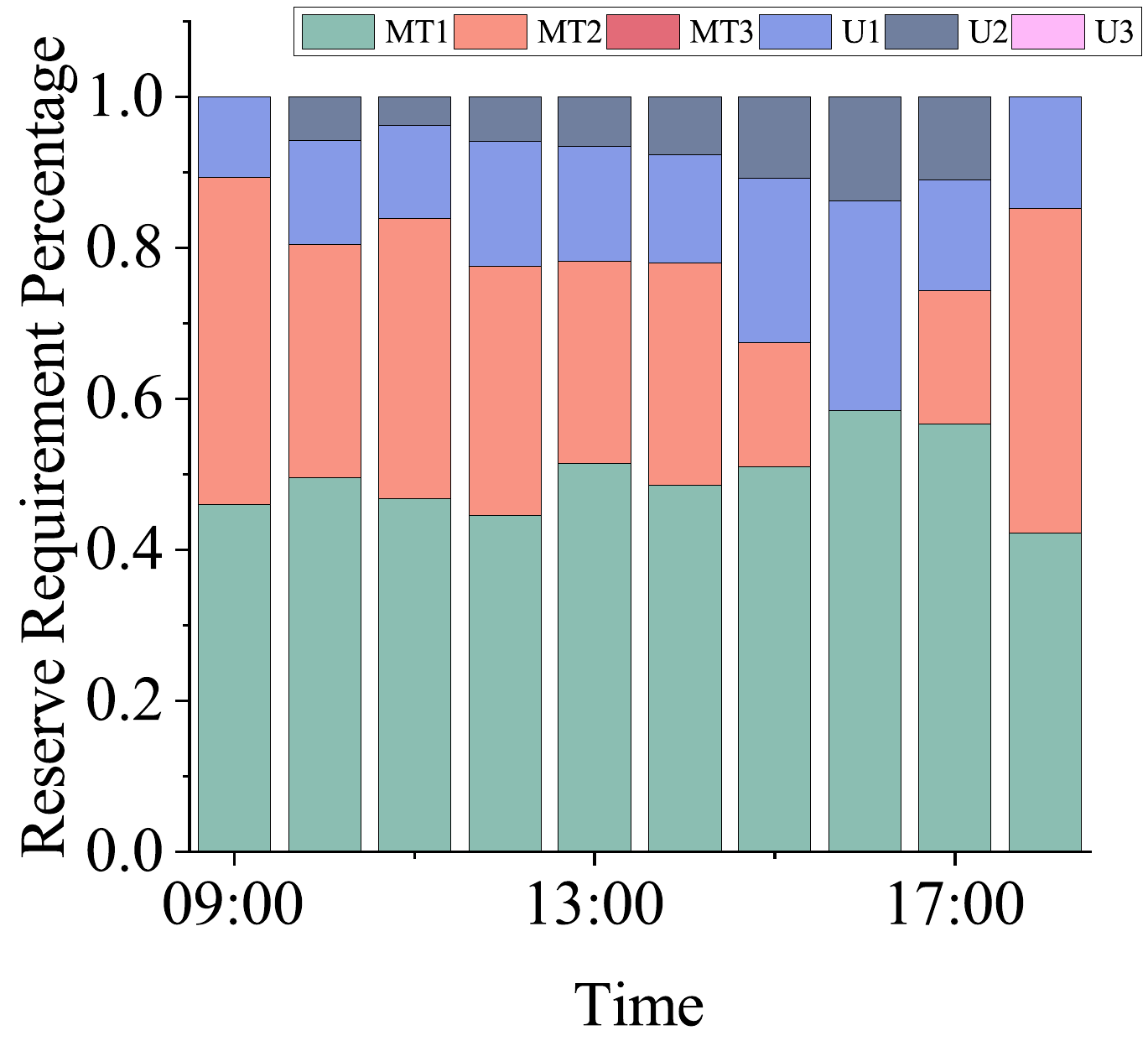}}
\caption{Uncertainty balance of each PV in the day-ahead market}
\label{fig4}
\vspace{-4mm}
\end{figure}

To achieve carbon allowance balance, PVs are required to purchase allowances from the market and users can decide whether to procure allowances to consume more energy or sell superfluous allowances to make profits based on their consumption profiles. The outcome of the carbon market is provided in Table \ref{tab3}. It is revealed that in this case, all users can sell allowances to both PVs and the community manager by adjusting their consumption behaviors. Besides, following KKT conditions, it is trivial to state that the price in the allowance sharing is the same as the selling price to the community manager.
\begin{table}[tp]
\begin{center}
%\small
\caption{The outcome of the carbon market}
\label{tab3}
\renewcommand{\arraystretch}{1.2}
\begin{tabular}{| c | c | c | c | c | c |}
\hline
Participants& U1 & U2 & U3 & PV1 & PV2\\
\hline
\multicolumn{6}{|c|}{Allowance Sharing}\\
\hline
Price \big[\$/\rm{kg}\big] & \multicolumn{5}{c|}{0.003}\\
\hline
Quantity \big[\rm{kg}\big] & 204.06 & 204.06 & 204.06 & -307.05 & -305.13\\
\hline
\multicolumn{6}{|c|}{Sold to the Community Manager}\\
\hline
Price \big[\$/\rm{kg}\big] & \multicolumn{5}{c|}{0.003}\\
\hline
Quantity \big[\rm{kg}\big] & 584.20 & 987.95 & 716.55 & / & /\\
\hline
\end{tabular}
\end{center}
\vspace{-6mm}
\end{table}

Next, the influences of the carbon allowance price and electricity price on the market outcomes are discussed. It is assumed that the carbon allowance price ranges from \$0.001/kg to \$0.006/kg and the electricity selling price is increased from \$0.04/kWh to \$0.08/kWh. The variations in social welfare, total allowances sold to the manager and total PV generations sold to the manager are displayed in Fig. 5. 
\begin{figure}[!t]
\centering
\subfloat[Social Welfare]{\includegraphics[width=0.45\textwidth]{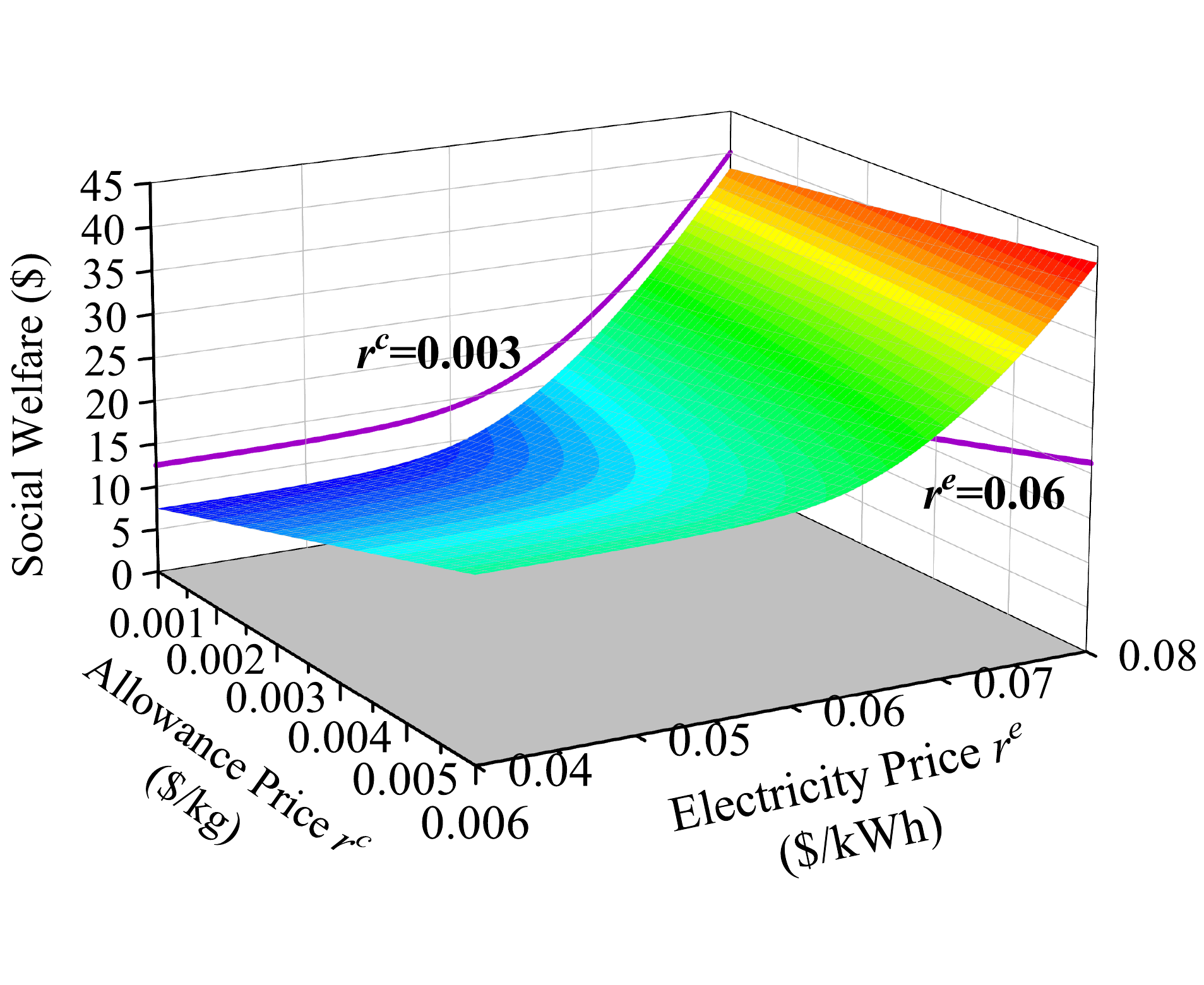}}\hfill
\subfloat[Total allowances sold to the manager]{\includegraphics[width=0.45\textwidth]{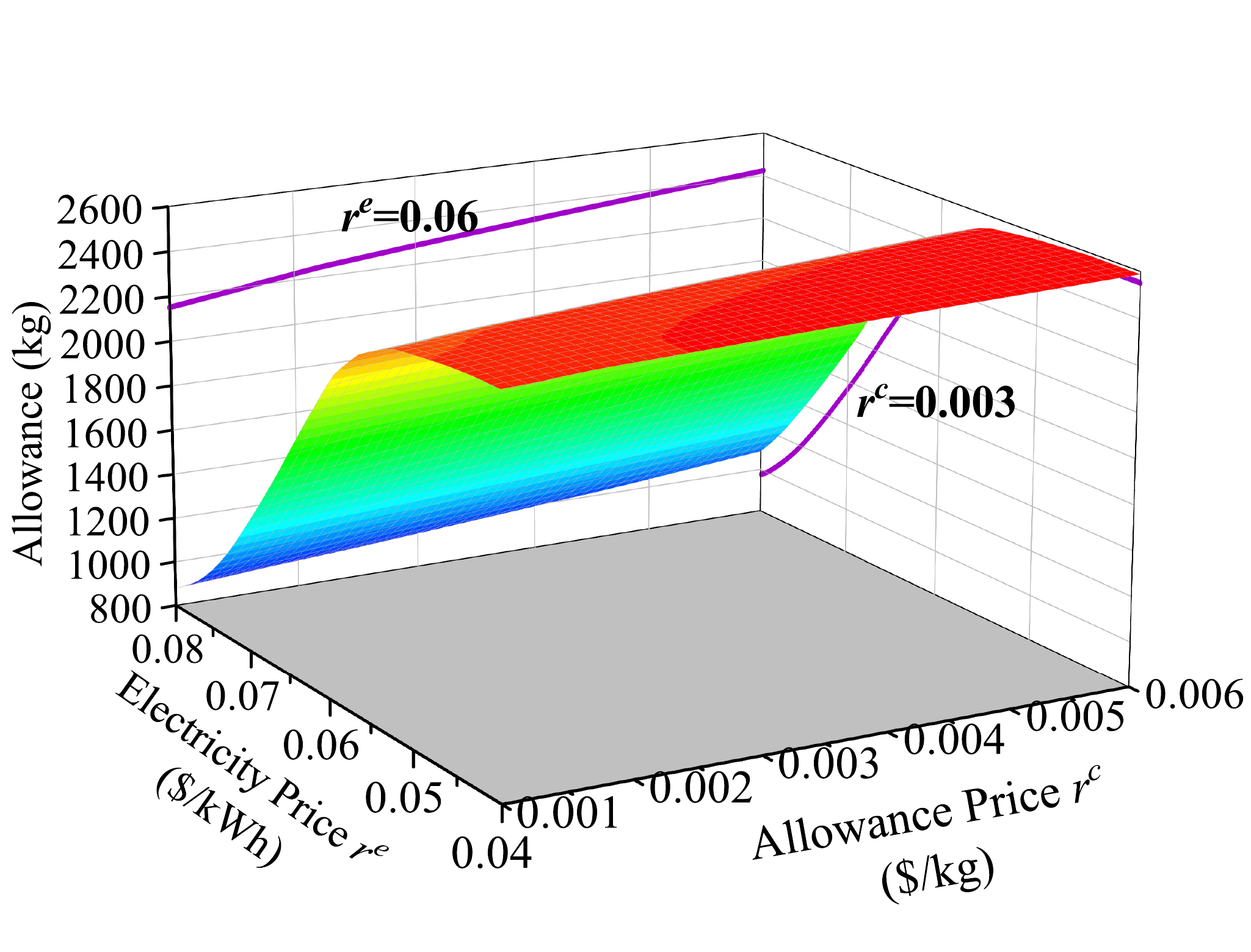}}\par
\subfloat[Total PV generations sold to the manager]{\includegraphics[width=0.45\textwidth]{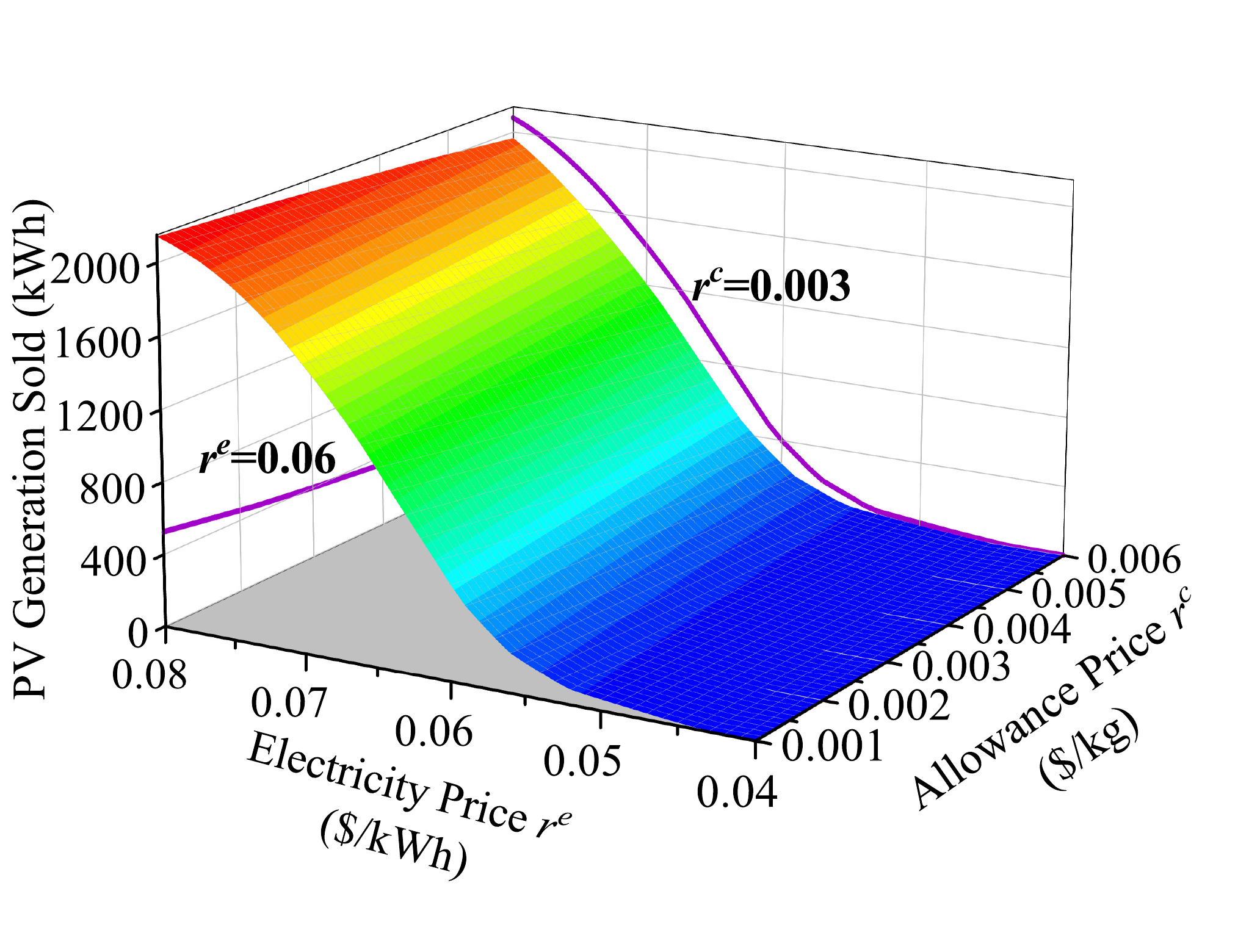}}
\caption{Social welfare, total allowances and PV generations sold to the manager as a function of electricity and carbon allowance price}
\label{fig5}
\vspace{-3mm}
\end{figure}

It can be observed from Fig. 5 (a) that the social welfare is improved along with the increment in electricity and allowance price since the whole community can sell allowances and PV generations at a more favorable price.
Fig. 5 (b)\&(c) reveal that the total allowances sold to the community manager are reduced with the increase of electricity price and the decrease of allowance price whereas the total PV generations sold to the manager soar. The decline in allowance price undermines the economic incentives for users to cut down daily consumption and to prefer green energy. Meanwhile, the increment in electricity price additionally stimulates the readiness of PVs to sell energy to the community manager instead of users. 

\subsection{The Effect of Carbon Allowance Sharing and Load Flexibility}
In this part, the effects of the carbon allowance sharing mechanism and load flexibility are demonstrated. To make a comparison, three different cases are considered here.

\textit{Case 1:} The joint market devised in this study is discussed, where carbon allowance sharing and load flexibility are both included.

\textit{Case 2:} The proposed joint market is considered, except for load flexibility.

\textit{Case 3:} The proposed joint market is adopted, while the carbon market model is modified according to \cite{carbon_blockchain}. In this case, the participants can only purchase/sell allowances to the operator at fixed prices and the allowance balancing constraint (9) is omitted. The selling price herein is still \$0.003/kg and the purchase price is set as \$0.009/kg.

The discrepancies in social welfare and total allowances held inside the community among the above three cases are presented in Table \ref{tab4}.
\begin{table}[!t]
\centering
%\small
\caption{Results in different cases}
\label{tab4}
\renewcommand{\arraystretch}{1.2}
\begin{tabular}{|c|c|c|} 
\hline
Cases & Social Welfare [\$] & \begin{tabular}[c]{@{}c@{}}Total Allowances Held Inside \\the Community [kg]\end{tabular}  \\ 
\hline
1     & 13.9858             & 3111.3                                                                                   \\ 
\hline
2     & 13.5043             & 3295.7                                                                                \\ 
\hline
3     & 10.4972                    & 3064.5                                                                                       \\
\hline
\end{tabular}
\end{table}

As can be vividly discerned from Table \ref{tab4}, the whole community has more demands for carbon allowances in case flexibility is not considered. This increase results from the need of PVs to balance uncertainty. In Case 2, the only access for PVs to balance uncertainty is purchasing upward reserve of MTs, which, therefore, inevitably precipitates more carbon emissions and demands for allowances. The decrease in the allowances sold to the community manager also lowers the social welfare. Hence, employing the flexibility of users occupies a crucial role in promoting social welfare and reducing carbon emissions.

In Case 3, the conspicuous shrinkage in social welfare is mainly due to the high cost of purchasing carbon allowances, which suppresses the need of PVs for allowances, and thereby saps the will of PVs to procure upward reserve of MTs. Compared with Case 1, it can be inferred that the carbon allowance sharing mechanism helps facilitate allowance trading among participants, generate a more affordable purchase price, and thus, improve total social welfare.  

\subsection{Convergence Analysis}
\begin{figure}
\centering
\subfloat[Primal residuals]{\includegraphics[width=0.8\linewidth]{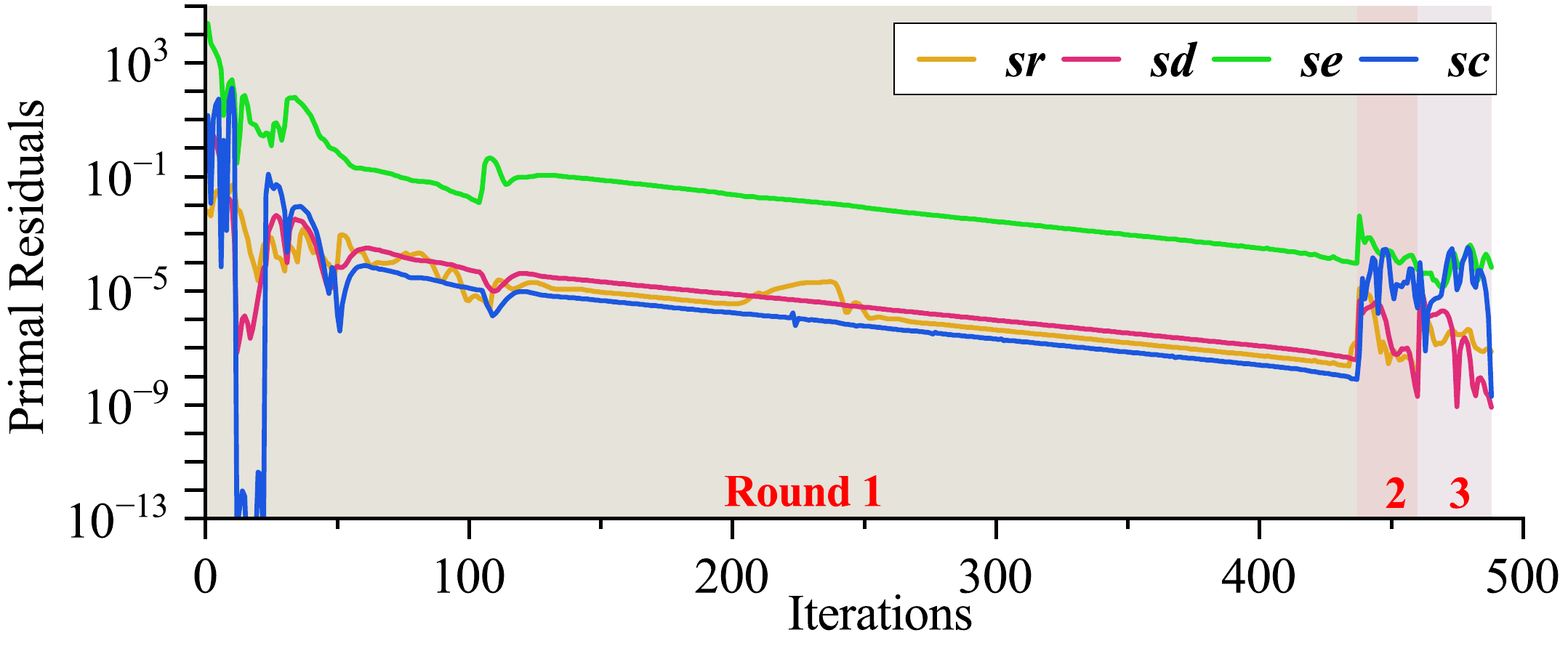}}

\subfloat[Dual residuals]{\includegraphics[width=0.8\linewidth]{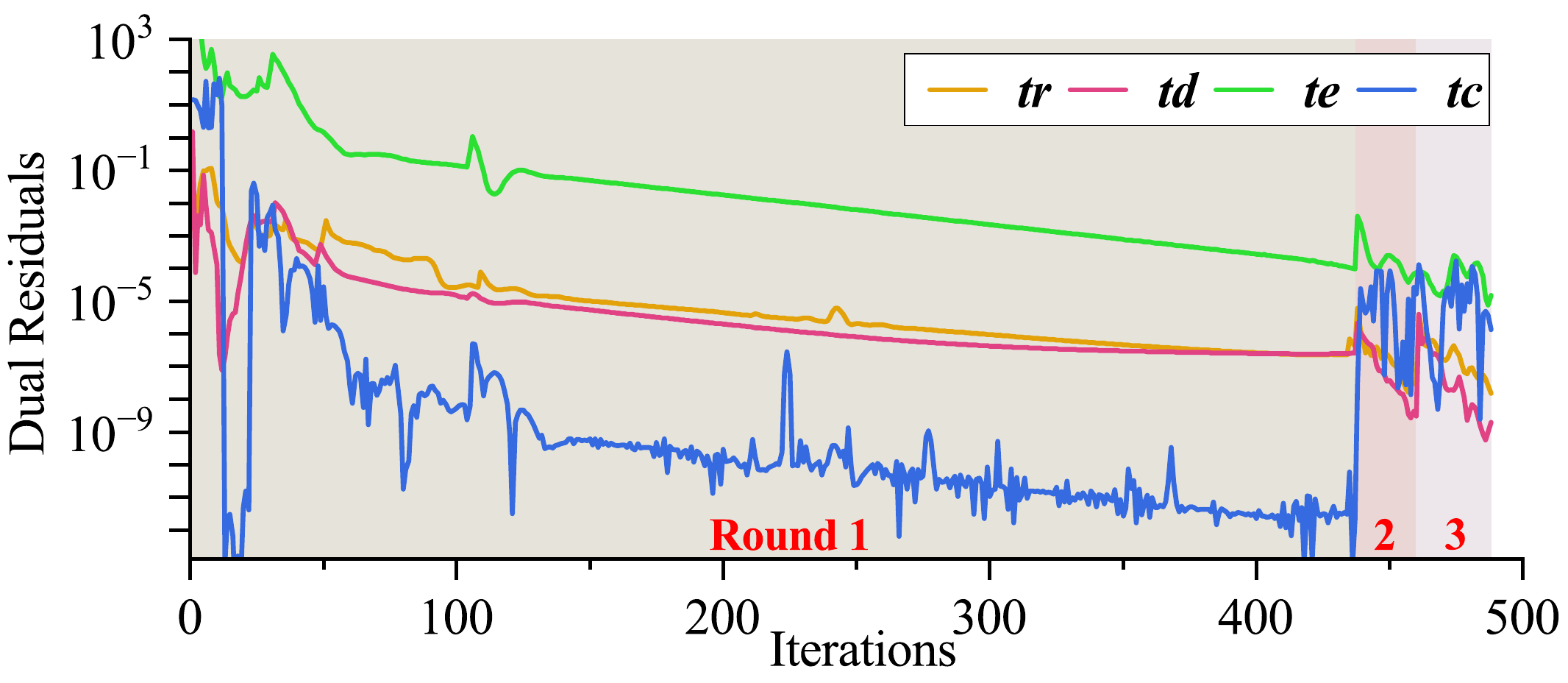}}
\caption{Primal and dual residuals during different rounds}
\label{fig6}
\vspace{-4mm}
\end{figure}

\begin{figure}
\centering
\includegraphics[width=0.8\linewidth]{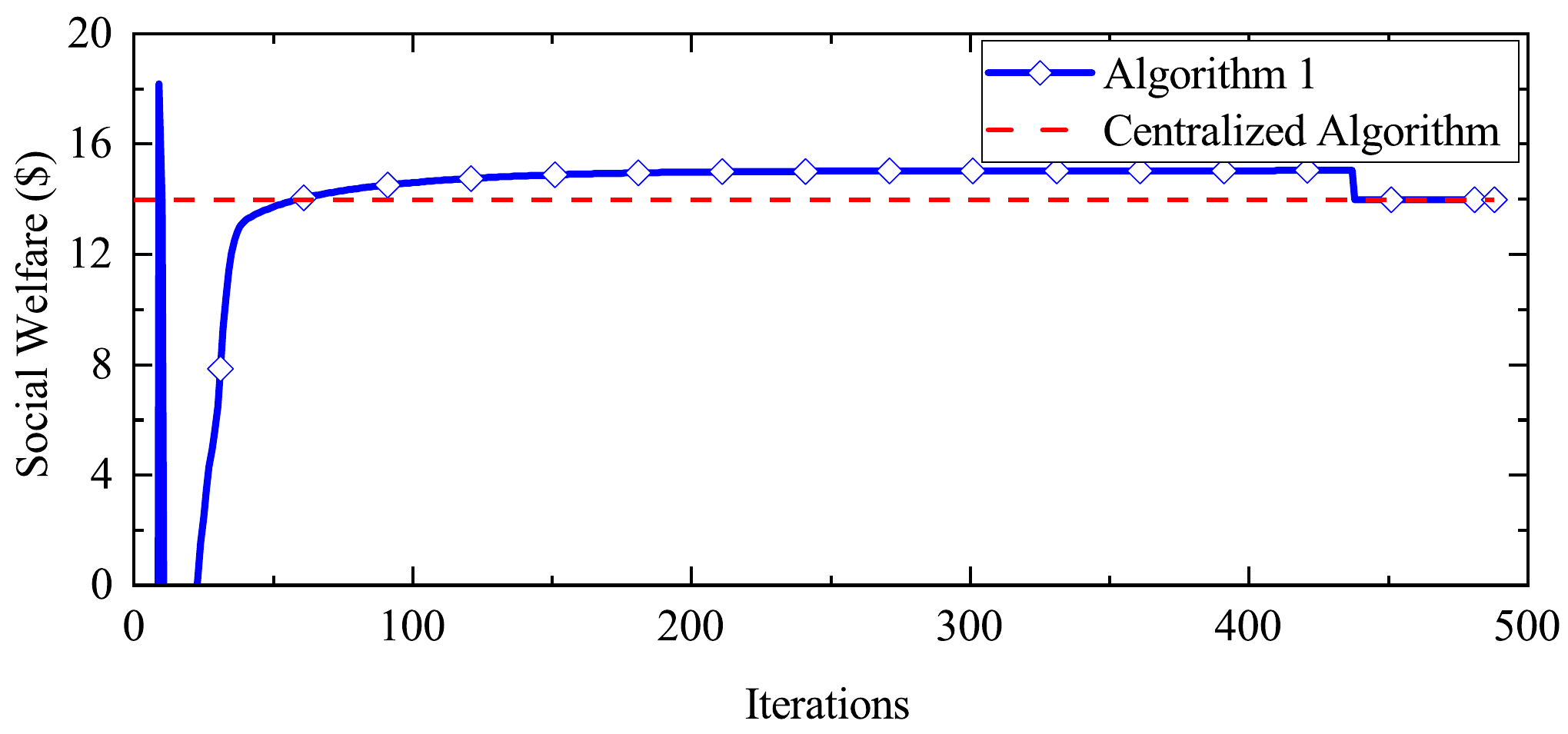}
\caption{Social welfare of the proposed algorithm and the centralized algorithm}
\label{fig7}
\vspace{-4mm}
\end{figure}
The evolution of the primal residuals and dual residuals through the proposed algorithm are plotted in Fig. 6. To accelerate the convergence speed, an adaptive penalty factor method as put forward in \cite[Chapter 3]{boyd2011distributed} is employed. As shown in Fig. 6, the proposed algorithm reaches stopping criteria within 3 bound contraction iterations (denoted as 'round') and a total of 488 ADMM iterations, verifying its convergence performance. Besides, the fluctuations in the social welfare are presented in Fig. 7. The optimal social welfare in the proposed algorithm is \$13.9858 and \$13.9869 in the centralized algorithm, which results in a negligible optimality gap of less than 0.01\%. This indicates that the proposed solution techniques can effectively cope with the decentralized trading in the energy community.

Next, we will herein illustrate the efficiency of the warm-start method mentioned in Section 3.3. Although the convergence of ADMM is guaranteed for any initialization points, a good starting point can dramatically reduce the iterations and computation complexity. In this study, it is reasonable to adopt the results obtained from the last round as an approximation of accurate ones since these results provide a lower bound to the original problem. Based on the fixed penalty factor version of ADMM, the total iterations required to converge under two different situations are provided in Table \ref{tab5}, one for results without a warm start and the other for results with a warm start. For the first round, both of the two methods involves 1969 ADMM iterations since they start from the same initialization points. However, for the second round, a transparent reduction in the number of iterations appears when the warm start is employed, demonstrating that the warm-start method contributes to the superior convergence performance. By comparison with Fig. 6, it is also found that a marked increase in total iterations occurs when adopting fixed penalty factor method, which verifies the superiority of the adaptive penalty factor method.
\begin{table}[!t]
\centering
%\small
\caption{Iteration numbers with/without warm start}
\label{tab5}
\begin{tabular}{|c|c|c|c|} 
\hline
\textbf{Iterations} & Round 1 & Round 2 & Total  \\ 
\hline
No Warm start       & 1969    & 1518    & 3487   \\ 
\hline
With Warm start     & 1969    & 33      & 2002   \\
\hline
\end{tabular}
\end{table} 

\section{Conclusion}
\label{5}
This paper proposes a joint day-ahead market paradigm where participants simultaneously trade energy, uncertainty, and carbon allowances. Moreover, this paper considers the possible excessive carbon emissions caused during uncertainty balance, which are non-negligible for the management of the carbon emission. Simulations have revealed several merits of the devised market: 1) The introduction of carbon allowance trading guarantees that the total carbon emission of the energy community does not exceed the prescribed limit; 2) Proper electricity price and allowance price are conducive to the local consumption of renewable energy and the reduction of carbon emissions; 3) The renewable agents can fully offset the uncertainty by procuring reserve from conventional generators and flexibility from users; 4) The proposed Relax-ADMM-Contraction loop is privacy-friendly, and can simultaneously yield trading quantities and trading prices. However, it should be noted that the Chebyshev approximation is generally far too conservative, and thereby may lead to the decline in the market efficiency. A further study should therefore concentrate on the methods to bypass the conservatism. 
\section*{Acknowledgements}
This work was supported in part by the National Key R \& D Program of China (No. 2020YFE0200400), and in part by the National Natural Science Foundation of China (No. 52177077). 
%\bibliographystyle{elsarticle-num} 
%\bibliography{main}

\appendix
\section{}
\nomenclature[A,01]{
\(\Omega_g\)}{Set of conventional generators}
\nomenclature[A,02]{
\(\Omega_u\)}{Set of users}
\nomenclature[A,03]{
\(\Omega_r\)}{Set of renewable agents}
\nomenclature[A,04]{
\(\xi_i^g\)}{Decision variable set of conventional generator \textit{i}}
\nomenclature[A,05]{
\(\xi_i^u\)}{Decision variable set of user \textit{i}}
\nomenclature[A,06]{
\(\xi_i^r\)}{Decision variable set of renewable agent \textit{i}}
\nomenclature[A,07]{
\(t\in T\)}{Index of time steps}

\nomenclature[C,01]{
\(\mu_i^t\)}{Expectation of \(\omega_i^{t-}\)}
\nomenclature[C,02]{
\(\delta_i^t\)}{Variance of \(\omega_i^{t-}\)}
\nomenclature[C,03]{
\(\Psi_i^0\)}{Initial carbon allowance allocated to user \textit{i}}
\nomenclature[C,04]{
\(\underline{p}_{u/g,i}^t\)}{Lower bound of user/conventional generator \textit{i} }
\nomenclature[C,05]{
\(\overline{p}_{u/g,i}^t\)}{Upper bound of user/conventional generator \textit{i} }
\nomenclature[C,06]{
\(P_{r,i}^t\)}{Forecast generation of renewable agent \textit{i} }
\nomenclature[C,07]{
\(\sigma_i\)}{Carbon intensity of conventional generator \textit{i} }

\nomenclature[V,01]{
\(Es_{ij}^t\)}{Energy quantity trade from seller \textit{j} to buyer \textit{i} at time \textit{t}}
\nomenclature[V,02]{
\(\alpha_{ij}^t\)}{Participation factor of conventional generator \textit{i}}
\nomenclature[V,03]{
\(\beta_{ij}^t\)}{Participation factor of user \textit{i}}
\nomenclature[V,04]{
\(c_i\)}{Carbon allowance trade quantity inside the community}
\nomenclature[V,05]{
\(c_i^s\)}{Carbon allowance quantity sold to the manager}
%\nomenclature[V,06]{
%\(\Psi_i\)}{Carbon allowance after trading}
\nomenclature[V,07]{
\(p_{u,i}^t\)}{Energy usage of user \textit{i} at time \textit{t}}
\nomenclature[V,08]{
\(p_{g,i}^t\)}{Output of conventional generator \textit{i} at time \textit{t}}
\nomenclature[V,09]{
\(p_{r,i}^t\)}{Energy traded in the community of renewable agent \textit{i} at time \textit{t}}
\setlength{\nomitemsep}{0.02cm}
\printnomenclature[1cm]




% \end{thebibliography}
\end{document}